\documentclass[twocolumn,showpacs,preprintnumbers,amsmath,amssymb]{revtex4}

\usepackage{graphicx}
\usepackage{dcolumn}
\usepackage{bm}

\begin{document}

\preprint{}

\title{Efficient Heating of Thin Cylindrical Targets\\by Broad Electromagnetic Beams II}

\author{Andrey Akhmeteli}
 \email{akhmeteli@aim.com}
\affiliation{%
23525 Arlington Ave. 120\\
Torrance, CA 90501, USA
}%


\date{November 17, 2006}

\begin{abstract}
In Part I (physics/0405091), it was shown that it is possible to achieve efficient heating of cylindrical targets by electromagnetic beams with transverse dimensions that are several orders of magnitude greater than those of the cylinder.

Part II contains derivation of the detailed conditions of efficient heating in the longitudinal geometry and establishes a broader domain  of parameters providing efficient heating in the transverse geometry.

One possible implementation using currently available technology is considered: a nanotube (length 1 cm, diameter 15 nm) may be heated in the transverse geometry by a 30 femtosecond pulse of a 50 GW Ti:Sapphire laser (wavelength 800 nm) to a temperature of 1 keV with heating efficiency of 16 per cent. The resulting energy density exceeds 100 MJ/cm3.

An exact formula in the form of a one-dimensional integral is obtained for the amount of energy absorbed in the cylinder heated by a gaussian electromagnetic beam in the longitudinal geometry.
\end{abstract}

\maketitle

\section{\label{sec:level11}Introduction}

In Part I (Ref.~\cite{Akhm10}), it was shown that it is possible to achieve efficient heating of cylindrical targets by electromagnetic beams with transverse dimensions that are several orders of magnitude greater than those of the cylinder.

Part II contains derivation of the detailed conditions of efficient heating in the longitudinal geometry and establishes a broader domain  of parameters providing efficient heating in the transverse geometry.

One possible implementation using currently available technology is considered: a nanotube (length 1 cm, diameter 15 nm) may be heated in the transverse geometry by a 30 femtosecond pulse of a 50 GW Ti:Sapphire laser (wavelength 800 nm) to a temperature of 1 keV with heating efficiency of 16 per cent. The resulting energy density exceeds 100 MJ/cm3.

An exact formula in the form of a one-dimensional integral is obtained for the amount of energy absorbed in the cylinder heated by a gaussian electromagnetic beam in the longitudinal geometry.

\maketitle
\section{\label{sec:level12}Domains of parameters providing high absorption efficiency in the longitudinal geometry}
In Part I of this work (Ref.~\cite{Akhm10}), the following asymptotic formula for absorption efficiency in the longitudinal geometry was derived:
\begin{eqnarray}\label{eq:700}
\eta\approx\frac{W^a}{W}=\textrm{Im} \frac{4}{p_2\frac{J_0(p_2)}{J_1(p_2)}\left(\frac{1}{p_2^2}+\frac{1}{2 a^2}\right)+\ln p_1}.
\end{eqnarray}
Here $\eta$ is the absorption efficiency, $W^{a}$ is the power absorbed in the cylinder, $W$ is the power in the incident beam, $a$ is the radius of the cylinder (a system of units is used where the wave vector in free space $k_{0}=\frac{\omega}{c}=1$), $p_1\approx\frac{2 a}{r_1}\ll 1$, $r_1$ is the beam waist radius defined by $e$-fold field intensity attenuation,
\begin{eqnarray}\label{eq:701}
p_2^2=(\varepsilon-1)a^2+p_1^2,
\end{eqnarray}
$\varepsilon$ is the complex permittivity of the cylinder. One can see that the refracted fields and, therefore, the absorption efficiency do not change when $p_2$ is substituted with $-p_2$. Therefore we shall assume that
\begin{eqnarray}\label{eq:701a}
p''=\textrm{Im}(p_2)>0
\end{eqnarray}
($p_2$ can only be real if $\varepsilon$ is real and, therefore, there is no absorption in the cylinder, so we do not consider this case).

It should also be noted that within the accuracy of the approximations used to derive Eq.~(\ref{eq:700}) it is possible to substitute $p_2$ in that equation with $\sqrt{(\varepsilon-1)}a$.

In this section, we are going to find the domains of parameters that provide high absorption efficiency in the asymptotic case $p_1\ll 1$, $l=|\ln(p_1)|\gg 1$. We consider absorption efficiency "high" if it is at least of the same order of magnitude as $\frac{1}{l}$. Although the value of the latter expression tends to zero as $p_1$ tends to zero, it decreases very slowly as $r_1$ increases and is quite high (typically tens of percent) for any realistic values of $r_1$. We are only interested in parameters that provide high efficiency when $p_1$ varies by an order of magnitude, as if absorption efficiency is only high for one value of $p_1$, it may be low for the relevant Gaussian beam, which is a linear combination of cylindrical waves with different values of $p_1$ (Eq.~(\ref{eq:700}) was derived using the one-wave approximation).

The right-hand side of Eq.~(\ref{eq:700}) strongly depends on three parameters: the real and imaginary parts of complex permittivity $\varepsilon$ and the radius of the cylinder $a$.
Analysis of Eq.~(\ref{eq:700}) may be broken down into several cases depending on the value of $p_2$. Essentially, depending on this value, we use four different approximations for the Bessel functions: for $p_2\alt 1$, for $p''_2\agt 1$, for $p_2$ in the vicinity of a positive zero of function $J_1(x)$, and, finally, for $p_2$ in the vicinity of a positive zero of function $J_0(x)$. Evidently, the domains of applicability of the relevant approximations partially overlap.

This analysis is rather cumbersome, but it has some elements common to all four major cases. These elements may be illustrated as follows.
If
\begin{eqnarray}\label{eq:701b}
p_2\frac{J_0(p_2)}{J_1(p_2)}\left(\frac{1}{p_2^2}+\frac{1}{2 a^2}\right)=X=X'-i X'',
\end{eqnarray}
where $X'$ and $X''$ are real, then
\begin{eqnarray}\label{eq:701c}
\eta\approx\textrm{Im} \frac{4}{X'-i X''-l}=\textrm{Im} \frac{4(X'+i X''-l)}{(X'-l)^2+ X''^2}=
\nonumber\\
=\frac{4 X''}{(X'-l)^2+ X''^2}.
\end{eqnarray}
For the purposes of illustration, let us treat $X'$ and $X''$ as independent parameters. The standard procedure of maximization over, say, $X''$ is complicated by the fact that $l$ is defined to an accuracy of unity (in the expansion of the fields of a Gaussian beam into cylindrical waves, values of $p_1$ differing by an order of magnitude make a significant contribution to the power of the incident beam -- see Eq.(22) of the first part of this work (Ref.~\cite{Akhm10})). As a result, we have to consider two cases: $|X'-l|\gg 1$ and $|X'-l|\alt 1$. If $|X'-l|\gg 1$, then $X''=|X'-l|$, $\eta=\frac{2}{|X'-l|}$ in the maximum, and the maximum is rather broad (its width at the level of 50\% of the maximum corresponds to variation of the argument by at least an order of magnitude). Obviously, $\eta\agt \frac{1}{l}$ if $|X'|\alt l$. On the other hand, if $|X'-l|\alt 1$, then $X''\sim 1$ and $\eta\sim 1$ in the maximum. Moreover, $\eta\agt \frac{1}{l}$ if
\begin{eqnarray}\label{eq:701d}
\frac{1}{l}\alt X''\alt l.
\end{eqnarray}
It is important to emphasize the difference between these two cases from the point of view of the subsequent analysis. For the analysis to cover all possible values of $p_2$, the approximations for the Bessel functions are used in relatively broad ranges. Therefore, for some values of $p_2$ the worst accuracy of the approximations for function $\frac{J_0(x)}{J_1(x)}$ may be, say, 30\%. Nevertheless, the essential formulae for the first case (e.g. $X''\sim |X'-l|$, $|X'|\alt l$,  $\eta\agt \frac{1}{l}$) still hold true in the sense that the optimal values calculated using the approximations will provide high absorption efficiency. On the other hand, this lack of accuracy does affect the essential formula for the second case: $|X'-l|\alt 1$, so high absorption efficiency will be provided only in the upper part of the wide range of Eq.~(\ref{eq:701d}), if the optimal values are calculated using the approximations. This defect of the analysis is somewhat mitigated by the fact that $l$ cannot realistically be much greater than 20. One may also question the value of asymptotic formulae derived using the approximation $l\gg 1$ (that means that the Gaussian beam is several orders of magnitude wider than the cylinder). The author believes, however, that these formulae will be a very useful guide to more realistic cases. In particular, it should be emphasized that for smaller $l$ it is generally easier to achieve high absorption efficiency, so the formulae show that the longitudinal geometry may have great practical value.
\subsection{\label{sec:level221}$|p_2|\alt 1$}

In this case we have
\begin{equation}\label{eq:702}
\frac{J_0(p_2)}{J_1(p_2)}\approx\frac{1}{\left(\frac{p_2}{2}\right)},
\end{equation}
thus
\begin{equation}\label{eq:703}
\eta\approx\textrm{Im} \frac{4}{\frac{2}{p_2^2}+\frac{1}{a^2}+\ln p_1}.
\end{equation}
As we assume that $p_1\ll|p_2|$, Eq.~(\ref{eq:701}) yields
\begin{equation}\label{eq:704}
\varepsilon\approx 1+\frac{p_2^2}{a^2},
\end{equation}
therefore,
\begin{eqnarray}\label{eq:705}
\eta\approx\textrm{Im} \frac{4}{\frac{2}{a^2(\varepsilon-1)}+\frac{1}{a^2}+\ln p_1}=
\nonumber\\
=\frac{4}{l}\textrm{Im} \frac{1}{\frac{1}{a^2l}\left(\frac{2}{(\varepsilon-1)}+1\right)-1}.
\end{eqnarray}
Let us define real values $x$, $y$, and $\varepsilon''$ by the following equalities:
\begin{equation}\label{eq:706}
x=\frac{1}{a^2 l}
\end{equation}
and
\begin{equation}\label{eq:707}
\varepsilon-1=y+i\varepsilon''.
\end{equation}
Then
\begin{eqnarray}\label{eq:708}
\eta\approx\frac{4}{l}\textrm{Im} \frac{\varepsilon-1}{x(\varepsilon-1)+2 x-(\varepsilon-1)}=
\nonumber\\
=\frac{4}{l}\textrm{Im} \frac{\varepsilon-1}{(x-1)(\varepsilon-1)+2 x}=
\nonumber\\
=\frac{4}{l}\textrm{Im} \frac{(y+i\varepsilon'')((y-i\varepsilon'')(x-1)+2 x)}{(y(x-1)+2 x)^2+\varepsilon''^2(x-1)^2}=
\nonumber\\
=\frac{4}{l}\frac{2 x\varepsilon''}{(y(x-1)+2 x)^2+\varepsilon''^2(x-1)^2}=
\nonumber\\
=\frac{4}{l}\frac{1}{(x-1)^2}\frac{2 x\varepsilon''}{\left(y+\frac{2 x}{x-1}\right)^2+\varepsilon''^2}.
\end{eqnarray}
As $p_1$ is defined with an "accuracy" of an order of magnitude, it may be roughly said that $l$ is defined with an accuracy of unity. Let us determine to what accuracy $\delta$ value $\frac{2 x}{x-1}$ is defined:
\begin{equation}\label{eq:709}
\frac{2 x}{x-1}=\frac{\frac{2}{a^2 l}}{\frac{1}{a^2 l}-1}=\frac{2}{1-a^2 l},
\end{equation}
so
\begin{eqnarray}\label{eq:710}
\delta=\Delta\left(\frac{2 x}{x-1}\right)=\Delta\left(\frac{2}{1-a^2 l}\right)\approx\frac{2 a^2\Delta l}{(1-a^2 l)^2}\sim
\nonumber\\
\sim\frac{2 a^2}{(1-a^2 l)^2}.
\end{eqnarray}
On the other hand,
\begin{equation}\label{eq:711}
x-1=\frac{1}{a^2 l}-1=\frac{1-a^2 l}{a^2 l},
\end{equation}
\begin{equation}\label{eq:712}
\frac{1}{(x-1)^2}=\frac{a^4 l^2}{(1-a^2 l)^2}.
\end{equation}
Thus, if $x\approx 1$, then $1\approx a^2 l$, $l\approx\frac{1}{a^2}$, value $1-a^2 l$ is defined with an accuracy of $a^2\Delta l\sim a^2\sim\frac{1}{l}$, $x$ is defined with an accuracy of $\frac{1}{a^2 l^2}\sim\frac{1}{l}$. Therefore, if $|x-1|\gg\frac{1}{l}$, then value $\frac{1}{|x-1|}$ is sufficiently well defined, and value $\frac{2 x}{x-1}$ is defined with an accuracy of
\begin{eqnarray}\label{eq:713}
\frac{2 a^2}{(1-a^2 l)^2}=\frac{1}{(x-1)^2}\frac{2}{a^2 l^2}=\frac{2 x}{(x-1)^2}\frac{1}{l}
\end{eqnarray}
(see Eqs.~(\ref{eq:710},\ref{eq:712})).

If, on the other hand, $|x-1|\sim\frac{1}{l}$, then
\begin{equation}\label{eq:714}
\left|\frac{2 x}{x-1}\right|\sim|2 x|l\sim2 l
\end{equation}
(as $x\approx 1$).
Let us consider this case.
\subsubsection{\label{sec:level311}$|x-1|\sim\frac{1}{l}$}
Eq.~(\ref{eq:708}) yields
\begin{equation}\label{eq:715}
\eta\sim\frac{4}{l}l^2\frac{2 \varepsilon''}{(y\pm 2 l)^2+\varepsilon''^2}.
\end{equation}
In the maximum
\begin{equation}\label{eq:716}
\varepsilon''\sim\max(|y|,2 l)
\end{equation}
and
\begin{equation}\label{eq:717}
\eta\sim\frac{4}{l}l^2\frac{1}{\max(|y|,2 l)}.
\end{equation}
Condition $\eta\agt\frac{1}{l}$ is met if $|y|\alt l^2$.
On the other hand, $|p_2|\alt 1$, or $|(\varepsilon-1)a^2|\alt 1$ (see Eq.~(\ref{eq:704})), thus
\begin{eqnarray}\label{eq:718}
|y|\alt\frac{1}{a^2}\sim l,
\nonumber\\
\varepsilon''\alt l.
\end{eqnarray}
Condition $\eta\agt\frac{1}{l}$ is met if $\varepsilon''\agt 1$.
Thus, we obtain the following conditions of high absorption efficiency in this case:
\begin{eqnarray}\label{eq:720}
|x-1|\sim \frac{1}{l},
\nonumber\\
\varepsilon''\sim l,
\nonumber\\
y\alt l.
\end{eqnarray}
\subsubsection{\label{sec:level312}$|x-1|\gg\frac{1}{l}$}
In this case, absorption efficiency (see Eq.~(\ref{eq:708})) has a maximum when
\begin{equation}\label{eq:721}
\varepsilon''=\max\left(\left|y+\frac{2 x}{x-1}\right|,\frac{2 x}{(x-1)^2 l}\right)
\end{equation}
(we have taken into account Eq.~(\ref{eq:713})).

Condition $\eta\agt\frac{1}{l}$ may only be met if
\begin{equation}\label{eq:722}
\left|y+\frac{2 x}{x-1}\right|\alt \frac{x}{(x-1)^2}.
\end{equation}
Let us consider the following cases.
\paragraph{\label{sec:level4121}$x\alt 1$.}

If $x\alt 1$, Eq.~(\ref{eq:722}) is equivalent to the following equation:
\begin{equation}\label{eq:723}
|y|\alt \frac{x}{(x-1)^2}.
\end{equation}
In fact, if Eq.~(\ref{eq:723}) is true, then
\begin{eqnarray}\label{eq:724}
\left|y+\frac{2 x}{x-1}\right|\leq |y|+\left|\frac{2 x}{x-1}\right|\alt
\nonumber\\
\alt\frac{x}{(x-1)^2}+|x-1|\frac{2 x}{(x-1)^2}\alt\frac{x}{(x-1)^2}.
\end{eqnarray}
If, on the other hand, Eq.~(\ref{eq:723}) is false, and $|y|\gg \frac{x}{(x-1)^2}$, then
\begin{eqnarray}\label{eq:725}
\left|y+\frac{2 x}{x-1}\right|\geq \left||y|-\left|\frac{2 x}{x-1}\right|\right|=
\nonumber\\
=\left||y|-2|x-1|\frac{x}{(x-1)^2}\right|\sim |y|\gg\frac{x}{(x-1)^2},
\end{eqnarray}
as $x\alt 1$ and $x>0$.

Thus, the conditions of high absorption efficiency are:
\begin{equation}\label{eq:726}
|x-1|\gg\frac{1}{l},
\end{equation}
\begin{equation}\label{eq:727}
x\alt 1,
\end{equation}
\begin{equation}\label{eq:728}
\varepsilon''\sim\max\left(\left|y+\frac{2 x}{x-1}\right|,\frac{2 x}{(x-1)^2 l}\right),
\end{equation}
plus Eq.~(\ref{eq:723}), and
\begin{equation}\label{eq:729}
|p_2|\alt 1.
\end{equation}
As
\begin{equation}\label{eq:730}
|(\varepsilon-1)a^2|\alt 1
\end{equation}
(see Eq.~(\ref{eq:704})), we have $|y|\alt\frac{1}{a^2}$, $\varepsilon''\alt\frac{1}{a^2}$, and, in view of Eq.~(\ref{eq:727}), $\frac{1}{a^2 l}\alt 1$.

Taking into account Eq.~(\ref{eq:726}), we obtain
\begin{equation}\label{eq:731}
\frac{2 x}{(x-1)^2 l}\ll\frac{2 x l^2}{l}=2 x l=\frac{2}{a^2},
\end{equation}
or
\begin{equation}\label{eq:732}
\frac{2 x}{(x-1)^2 l}\ll\frac{1}{a^2},
\end{equation}
and
\begin{equation}\label{eq:733}
\left|\frac{2 x}{(x-1)}\right|\ll 2 x l=\frac{2}{a^2},
\end{equation}
or
\begin{equation}\label{eq:734}
\left|\frac{2 x}{(x-1)}\right|\ll\frac{1}{a^2}.
\end{equation}
Thus, if $|y|\alt\frac{1}{a^2}$, then, in view of Eq.~(\ref{eq:728}), condition $|p_2|\alt 1$ is satisfied.
\paragraph{\label{sec:level4122}$x\gg 1$.}

In this case we have Eq.~(\ref{eq:728}) again. As $x\gg 1$, we also obtain
\begin{equation}\label{eq:735}
\frac{2 x}{(x-1)^2 l}\sim \frac{2}{x l}.
\end{equation}
From Eq.~(\ref{eq:722}) we have
\begin{equation}\label{eq:736}
\left|y+\frac{2 x}{x-1}\right|\alt \frac{x}{(x-1)^2}\sim\frac{1}{x},
\end{equation}
\begin{equation}\label{eq:737}
\left|y+\frac{2 x-2+2}{x-1}\right|\alt\frac{1}{x},
\end{equation}
\begin{equation}\label{eq:738}
|y+2|\alt\frac{1}{x}.
\end{equation}
Eq.~(\ref{eq:728}) may be rewritten in this case as
\begin{equation}\label{eq:739}
\varepsilon''=\max\left(\left|y+\frac{2 x}{x-1}\right|,\frac{2}{x l}\right).
\end{equation}
Eqs.~(\ref{eq:738}-\ref{eq:739}) imply $|p_2|\ll 1$. In fact, $x\gg 1$, therefore $\frac{1}{a^2}\gg l$; on the other hand, from Eq.~(\ref{eq:738}) we obtain $|y|\sim 1$, so $|y|\ll\frac{1}{a^2}$. On the other hand, $|\frac{2 x}{x-1}|\sim 1$, $\frac{2}{x l}\ll 1$, so Eq.~(\ref{eq:739}) yields $\varepsilon''\ll\frac{1}{a^2}$.

It is also necessary to check that in the expression for $D$ (Eq.~(136) of Part I of this work (Ref.~\cite{Akhm10})), the term with $\frac{1}{p_1^2}$ is at least of the same order of magnitude as the term with $p_1^0$.

The coefficient of $\frac{1}{p_1^2}$ in $D$ equals
\begin{equation}\label{eq:740}
-\frac{2}{p_2^2}-\frac{1}{a^2}-2 f_3-f_2\frac{p_2^2}{a^2},
\end{equation}
where
\begin{equation}\label{eq:741}
f_2=\frac{1}{p_2}\frac{J'_1(p_2)}{J_1(p_2)},
\end{equation}
\begin{equation}\label{eq:741a}
f_3=-l+f_2.
\end{equation}
Eqs.~(\ref{eq:738}-\ref{eq:739}) yield
\begin{equation}\label{eq:742}
|\varepsilon+1|\alt\frac{1}{x},
\end{equation}
where $x=\frac{1}{a^2 l}$. We may express $\varepsilon$ in the following form:
\begin{equation}\label{eq:743}
\varepsilon=-1+a^2 l_\alpha,
\end{equation}
where $l_\alpha=l \alpha'$, $|\alpha'|\alt 1$. Then
\begin{equation}\label{eq:744}
\varepsilon-1=-2+a^2 l_\alpha,
\end{equation}
Eq.~(\ref{eq:704}) yields
\begin{equation}\label{eq:746}
\varepsilon-1\approx\frac{p_2^2}{a^2},
\end{equation}
therefore
\begin{eqnarray}\label{eq:747}
p_2\approx a\sqrt{\varepsilon-1}=a\sqrt{-2+a^2 l_\alpha}=
\nonumber\\
=a\sqrt{-2}\sqrt{1-\frac{a^2 l_\alpha}{2}}\approx a\sqrt{2}i\left(1-\frac{a^2 l_\alpha}{4}\right),
\end{eqnarray}
as $|\alpha'|\alt 1$, $x=\frac{1}{a^2 l}\gg1$, and $|p_2|\ll 1$.

Let us obtain an approximation for $f_2$, taking into account that $|p_2|\ll 1$:
\begin{eqnarray}\label{eq:748}
f_2=\frac{1}{p_2}\frac{J'_1(p_2)}{J_1(p_2)}\approx\frac{1}{p_2}\frac{\frac{1}{2}}{\frac{1}{2}p_2}=
\nonumber\\
=\frac{1}{p_2^2}\approx\frac{1}{-2 a^2\left(1-\frac{a^2 l_\alpha}{2}\right)}\approx
\nonumber\\
\approx-\frac{1}{2 a^2}\left(1+\frac{a^2 l_\alpha}{2}\right)=-\frac{1}{2 a^2}-\frac{l_\alpha}{4}.
\end{eqnarray}
We have taken into account that, e.g., substitution of $J'_1(p_2)$ by $\frac{1}{2}$ introduces an error equivalent to introduction of a factor $(1+\alpha''p_2^2)$, where $\alpha''\sim1$ and $|p_2|\ll 1$.

In view of Eq.~(\ref{eq:741a}), the coefficient of $\frac{1}{p_1^2}$ in $D$ equals:
\begin{eqnarray}\label{eq:748b}
-\frac{2}{-2 a^2\left(1-\frac{a^2 l_\alpha}{2}\right)}-\frac{1}{a^2}+2 l-2\left(-\frac{1}{2 a^2}-\frac{l_\alpha}{4}\right)-
\nonumber\\
-\left(-\frac{1}{2 a^2}-\frac{l_\alpha}{4}\right)\frac{-2 a^2\left(1-\frac{a^2 l_\alpha}{2}\right)}{a^2}\approx
\nonumber\\
\approx\frac{1}{a^2}\left(1+\frac{a^2 l_\alpha}{2}\right)-\frac{1}{a^2}+2 l+2\left(\frac{1}{2 a^2}+\frac{l_\alpha}{4}\right)-
\nonumber\\
-2\left(\frac{1}{2 a^2}+\frac{l_\alpha}{4}\right)\left(1-\frac{a^2 l_\alpha}{2}\right)=\frac{1}{a^2} +\frac{l_\alpha}{2}-
\nonumber\\
-\frac{1}{a^2}+2 l+\frac{1}{a^2}+\frac{l_\alpha}{2}-\frac{1}{a^2}-\frac{l_\alpha}{2}+\frac{l_\alpha}{2}+\frac{a^2l_\alpha^2}{4}\approx
\nonumber\\
\approx l_\alpha+2 l.
\end{eqnarray}
This coefficient is typically of the same order of magnitude as $l$ ($\sim l$). We may assume that it is $\agt 1$, as $l$ is defined with an accuracy of unity.

The coefficient of $p_2^0$ in $D$ equals
\begin{eqnarray}\label{eq:753}
\frac{1}{p_2^4}+\frac{2}{a^2 p_2^2}-f_3^2+\frac{f_2}{a^2}-f_3 f_2 \frac{p_2^2}{a^2}\approx
\nonumber\\
\approx\frac{1}{4 a^4}-\frac{1}{a^4}-\frac{1}{4 a^4}-\frac{1}{2 a^4}+\frac{2}{4 a^4}\approx-\frac{1}{a^4}.
\end{eqnarray}
Therefore, a sufficient condition ensuring that the term with $\frac{1}{p_2^2}$ in $D$ is at least of the same order of magnitude as the term with $p_2^0$ may be written as follows:

\begin{equation}\label{eq:754}
\frac{1}{p_1^2}\agt\frac{1}{a^4},
\end{equation}

\subsection{\label{sec:level222}$|p_2|\agt 1$}
Let us consider the following case:
\subsubsection{\label{sec:level321}$p''\agt 1$}
The approximation of the Bessel functions for an argument with a large magnitude may be used:
\begin{equation}\label{eq:756}
J_0(p_2)\approx\sqrt{\frac{2}{\pi p_2}}\cos(p_2-\frac{\pi}{4}),
\end{equation}
\begin{equation}\label{eq:757}
J_1(p_2)\approx\sqrt{\frac{2}{\pi p_2}}\cos(p_2-\frac{\pi}{2}-\frac{\pi}{4}),
\end{equation}
\begin{eqnarray}\label{eq:758}
\frac{J_0(p_2)}{J_1(p_2)}\approx\frac{\cos(p_2-\frac{\pi}{4})}{\cos(p_2-\frac{3\pi}{4})}\approx -\frac{\sin(p_2-\frac{3\pi}{4})}{\cos(p_2-\frac{3\pi}{4})}=
\nonumber\\
=-\tan\left(p_2-\frac{3 \pi}{4}\right)=-\frac{\tan(z)+\tan(i p'')}{1-\tan(z)\tan(i p'')},
\end{eqnarray}
where $z$ and $p''$ are real, and
\begin{equation}\label{eq:759}
z+i p''=p_2-\frac{3\pi}{4}.
\end{equation}
We obtain
\begin{eqnarray}\label{eq:760}
\tan(i p'')=\frac{\sin(i p'')}{\cos(i p'')}=\frac{\frac{1}{2 i}(\exp(-p'')-\exp(p''))}{\frac{1}{2}(\exp(-p'')+\exp(p''))}\approx
\nonumber\\
\approx-\frac{1}{i}=i,
\end{eqnarray}
\begin{equation}\label{eq:761}
-\frac{\tan(z)+i}{1-i\tan(z)}=-i,
\end{equation}
and
\begin{equation}\label{eq:762}
\frac{J_0(p_2)}{J_1(p_2)}\approx-i,
\end{equation}
\begin{eqnarray}\label{eq:763}
\eta\approx\textrm{Im} \frac{4}{-i p_2\left(\frac{1}{p_2^2}+\frac{1}{2 a^2}\right)+\ln p_1}=
\nonumber\\
=\textrm{Im} \frac{4}{-\frac{i}{p_2}-\frac{i p_2}{2 a^2}+\ln p_1}.
\end{eqnarray}
Let $p'=\textrm{Re}(p_2)$, so $p_2=p'+i p''$. Then
\begin{equation}\label{eq:764}
-\frac{i}{p_2}=-\frac{i}{p'+i p''}=\frac{-i(p'-i p'')}{p'^2+p''^2}=\frac{-ip'- p''}{p'^2+p''^2},
\end{equation}
\begin{equation}\label{eq:765}
-\frac{i p_2}{2 a^2}=-\frac{i (p'+i p'')}{2 a^2}=\frac{-i p'+p''}{2 a^2},
\end{equation}
thus,
\begin{eqnarray}\label{eq:766}
\eta\approx\textrm{Im} \frac{4}{\frac{-ip'- p''}{p'^2+p''^2}+\frac{-i p'+p''}{2 a^2}+\ln p_1}=
\nonumber\\
=\textrm{Im} \frac{4}{-\frac{ p''}{p'^2+p''^2}+\frac{p''}{2 a^2}+\ln p_1-i\left(\frac{ p'}{p'^2+p''^2}+\frac{p'}{2 a^2}\right)}.
\end{eqnarray}
Let us prove that
\begin{equation}\label{eq:767}
\left|\frac{p'}{p'^2+p''^2}\right|\alt 1.
\end{equation}
In fact, if $|p'|\geq p''$, then
\begin{eqnarray}\label{eq:768}
\left|\frac{p'}{p'^2+p''^2}\right|\sim\left|\frac{p'}{p'^2}\right|=\left|\frac{1}{p'}\right|\leq\frac{1}{p''}\alt 1,
\end{eqnarray}
on the other hand, if $|p'|<p''$, then
\begin{eqnarray}\label{eq:769}
\left|\frac{p'}{p'^2+p''^2}\right|\sim\frac{|p'|}{p''^2}<\frac{1}{p''}\alt 1.
\end{eqnarray}
Similarly,
\begin{equation}\label{eq:770}
\frac{p''}{p'^2+p''^2}\alt 1.
\end{equation}
In fact, if $|p'|\geq p''$, then
\begin{eqnarray}\label{eq:771}
\frac{p''}{p'^2+p''^2}\sim\frac{p''}{p'^2}\leq\frac{1}{|p'|}\leq\frac{1}{p''}\alt 1,
\end{eqnarray}
on the other hand, if $|p'|<p''$, then
\begin{eqnarray}\label{eq:772}
\frac{p''}{p'^2+p''^2}\sim\frac{p''}{p''^2}=\frac{1}{p''}\alt 1.
\end{eqnarray}
Therefore, in Eq.~(\ref{eq:766}), we may neglect value $\frac{p''}{p'^2+p''^2}$ in comparison with $|\ln(p_1)|$, as the latter is defined with an accuracy of unity. We may also neglect value $\frac{p'}{p'^2+p''^2}$ there for the following reason. Let us assume that
\begin{equation}\label{eq:773}
\eta\approx\textrm{Im} \frac{4}{-l+i\Delta_1},
\end{equation}
where $l=|\ln(p_1)|\gg 1$, $|\Delta_1|\alt 1$, and $\Delta_1$ is real. Then
\begin{equation}\label{eq:774}
\eta\approx\textrm{Im} \frac{4 (-l-i\Delta_1)}{l^2+\Delta_1^2}\sim\frac{-4\Delta_1}{l^2};
\end{equation}
we assume that this is too little (we are interested in values $\eta\agt \frac{1}{l}$).

While the magnitude of the real part of the denominator in Eq.~(\ref{eq:766}) may be much smaller than $l$, that would imply
\begin{equation}\label{eq:775}
\frac{p''}{2 a^2}\sim l\gg 1\agt\frac{p''}{p'^2+p''^2},
\end{equation}
therefore,
\begin{equation}\label{eq:776}
\left|\frac{p'}{2 a^2}\right|\gg\frac{|p'|}{p'^2+p''^2},
\end{equation}
so in this case we also may neglect $\frac{p'}{p'^2+p''^2}$ in Eq.~(\ref{eq:766}).

Therefore, we obtain
\begin{eqnarray}\label{eq:777}
\eta\approx\textrm{Im}\frac{4}{\frac{p''}{2 a^2}+\ln p_1-i\frac{p'}{2 a^2}}.
\end{eqnarray}
Let us introduce the following notation: $\alpha=\frac{p'}{2 a^2}$, $\beta=\frac{p''}{2 a^2}$. Again, $l=|\ln(p_1)|$. As $p''\agt 1$,
\begin{eqnarray}\label{eq:777a}
\beta\agt\frac{1}{2 a^2}.
\end{eqnarray}
Then
\begin{eqnarray}\label{eq:778}
\eta\approx\textrm{Im}\frac{4}{\beta-l-i\alpha}=\textrm{Im}\frac{4(\beta-l+i\alpha)}{(\beta-l)^2+\alpha^2}=
\nonumber\\
=\frac{4\alpha}{(\beta-l)^2+\alpha^2}.
\end{eqnarray}
In view of Eq.~(\ref{eq:704}), we obtain
\begin{eqnarray}\label{eq:779}
\varepsilon-1=y+i\varepsilon''\approx\frac{p_2^2}{a^2}=\frac{(p'+i p'')^2}{a^2}=
\nonumber\\
=\frac{p'^2-p''^2+2 i p'p''}{4 a^4}4 a^2=4 a^2(\alpha^2-\beta^2+2 i\alpha\beta).
\end{eqnarray}
Evidently, $\alpha\beta>0$, otherwise $\varepsilon''\leq 0$, and there is no absorption. From the definitions of $\alpha$ and $\beta$ and Eq.~(\ref{eq:701a}) we then obtain $\alpha>0$ and $\beta>0$.
High absorption efficiency may only be achieved if
\begin{equation}\label{eq:780}
\beta\alt l,
\end{equation}
as otherwise ($\beta\gg l$)
\begin{eqnarray}\label{eq:781}
\eta\approx\frac{4\alpha}{(\beta-l)^2+\alpha^2}\sim\frac{4\alpha}{\beta^2+\alpha^2},
\end{eqnarray}
and if $\alpha\geq\beta$, then
\begin{eqnarray}\label{eq:782}
\eta\sim \frac{4\alpha}{\alpha^2}=\frac{4}{\alpha}\leq\frac{4}{\beta}\ll \frac{1}{l},
\end{eqnarray}
whereas if $\alpha < \beta$ ,then
\begin{eqnarray}\label{eq:783}
\eta\sim \frac{4\alpha}{\beta^2}<\frac{4\beta}{\beta^2}=\frac{4}{\beta}\ll \frac{1}{l}.
\end{eqnarray}
Similarly,
\begin{equation}\label{eq:784}
\alpha\alt l,
\end{equation}
as otherwise ($\alpha\gg l$)
\begin{eqnarray}\label{eq:785}
\eta\approx\frac{4\alpha}{(\beta-l)^2+\alpha^2}\leq\frac{4\alpha}{\alpha^2}=\frac{4}{\alpha}\ll \frac{1}{l}.
\end{eqnarray}
As $l$ is defined with an accuracy of unity, let us consider the following two cases (as $l$ cannot realistically be much greater than 20, we shall commit a sin against mathematics and assume for the sake of simplicity that if $|\beta-l|\ll l$, then $|\beta-l|\alt 1$).
\paragraph{\label{sec:level4211}$|\beta-l|\sim 1$}
In this case
\begin{equation}\label{eq:786}
\frac{1}{l}\alt\alpha\alt l,
\end{equation}
as otherwise $\eta\ll \frac{1}{l}$. We may assume that $\frac{1}{l}\alt\alpha\alt 1$ (the combinations of $\alpha$ and $\beta$ that we thus omit ($|\beta-l|\sim 1$, $\alpha\sim l$) are considered together with the next case).

Let us introduce the following notation:
\begin{equation}\label{eq:786a}
u=\alpha^2-\beta^2,
\end{equation}
\begin{equation}\label{eq:786b}
v=2\alpha\beta.
\end{equation}
Then we obtain
\begin{equation}\label{eq:786c}
|u+l^2|\alt l,
\end{equation}
as $\beta=l+\Delta_\beta$, where $\Delta_\beta\alt1$, so $\beta^2=l^2+2 l \Delta_\beta+\Delta_\beta^2$. We also obtain
\begin{equation}\label{eq:786d}
2\alt v\alt 2 l.
\end{equation}
As $p''\agt 1$, then $\beta=\frac{p''}{2 a^2}\agt\frac{1}{2 a^2}$, so
\begin{equation}\label{eq:786e}
l\agt\frac{1}{2 a^2}.
\end{equation}
In view of the results for the next case (see below), we may write $2\alt v\alt 2 l^2$. Thus we obtain the following conditions for $y$ and $\varepsilon''$ (cf. Eqs.~(\ref{eq:779},\ref{eq:786a}-\ref{eq:786b})):
\begin{equation}\label{eq:786f}
|y+4 a^2 l^2|\alt 4 a^2 l,
\end{equation}
\begin{equation}\label{eq:786g}
8 a^2 \alt\varepsilon''\alt 8 a^2 l^2,
\end{equation}
\begin{equation}\label{eq:786h}
l\agt\frac{1}{2 a^2}.
\end{equation}
\paragraph{\label{sec:level4212}$|\beta-l|\sim l$.}
In this case
\begin{equation}\label{eq:787}
\alpha \sim l,
\end{equation}
 as otherwise $\eta\sim\frac{4 \alpha}{l^2}\ll\frac{1}{l}$ (cf. Eq.~(\ref{eq:784})). We may assume that in this case $\alpha\sim l$ and $\beta\alt l$, as the combinations of $\alpha$ and $\beta$ that we thus add ($|\beta-l|\sim 1$, $\alpha\sim l$) were omitted in the previous case. Let us further divide this case into the following two cases:
\subparagraph{\label{sec:level52121}$\alpha\sim l$, $\beta\sim l$.}

In view of Eqs.~(\ref{eq:777a},\ref{eq:786a}-\ref{eq:786b}), we obtain: $|u|\alt l^2$, $v\sim 2 l^2$, $l\agt\frac{1}{2 a^2}$. Thus, we have the following conditions for $y$ and $\varepsilon''$:
\begin{equation}\label{eq:790}
|y|\alt 4 a^2 l^2,
\end{equation}
\begin{equation}\label{eq:791}
\varepsilon''\sim 8 a^2 l^2,
\end{equation}
\begin{equation}\label{eq:792}
l\agt\frac{1}{2 a^2}.
\end{equation}
\subparagraph{\label{sec:level52122}$\alpha\sim l$, $\beta\ll l$.}
In view of Eq.~(\ref{eq:777a}), we obtain $u\sim l^2$, $\frac{l}{a^2}\alt v\ll 2 l^2$. In view of the conditions for the previous case, we may write for this case: $u\sim l^2$, $\frac{l}{a^2}\alt v\alt 2 l^2$. We obtain for $y$ and $\varepsilon''$:
\begin{equation}\label{eq:793}
y\sim 4 a^2 l^2,
\end{equation}
\begin{equation}\label{eq:794}
4 l\alt\varepsilon''\alt 8 a^2 l^2.
\end{equation}
\subsubsection{\label{sec:level322}Parameter $p_2$ in the vicinity of a zero of $J_1(x)$}
Let us now consider the case where $p_2=p_0+\Delta p$, $p_0$ being a nontrivial zero of function $J_1(x)$, so $p_0$ is real and positive (negative zeros are not relevant in view of Eqs.~(\ref{eq:701}-\ref{eq:701a})), and $|\Delta p|\alt 1$. For all $x$ we have
\begin{equation}\label{eq:795}
J'_0(x)=-J_1(x),
\end{equation}
\begin{equation}\label{eq:796}
J_0(x)=\frac{J_1(x)}{x}+J'_1(x),
\end{equation}
\begin{equation}\label{eq:797}
-J_1(x)=J'_0(x)=\frac{J'_1(x)}{x}-\frac{J_1(x)}{x^2}+J''_1(x),
\end{equation}
so
\begin{equation}\label{eq:798}
J_1(p_0)=0,
\end{equation}
\begin{equation}\label{eq:799}
J'_1(p_0)=J_0(p_0),
\end{equation}
\begin{equation}\label{eq:800}
J''_1(p_0)=-\frac{J_0(p_0)}{p_0}.
\end{equation}
Thus,
\begin{eqnarray}\label{eq:801}
\frac{J_0(p_0+\Delta p)}{J_1(p_0+\Delta p)}\approx
\nonumber\\
\approx\frac{J_0(p_0)+\Delta p J'_0(p_0)}{J_1(p_0)+\Delta p J'_1(p_0)+\frac{1}{2}(\Delta p)^2 J''_1(p_0)}=
\nonumber\\
=\frac{J_0(p_0)}{\Delta p J_0(p_0)-\frac{1}{2}(\Delta p)^2\frac{ J_0(p_0)}{p_0}}=
\nonumber\\
=\frac{1}{\Delta p}\left(\frac{1}{1-\frac{1}{2}\frac{\Delta p}{p_0}}\right)\approx\frac{1}{\Delta p}\left(1+\frac{1}{2}\frac{\Delta p}{p_0}\right)
\end{eqnarray}
(we left more terms in the expansion of the denominator as the first term there is zero).

Let us rewrite Eq.~(\ref{eq:700}) in the following form:
\begin{equation}\label{eq:802}
\eta\approx\textrm{Im} \frac{4}{\frac{J_0(p_2)}{J_1(p_2)}\left(\frac{1}{p_2}+\frac{p_2}{2 a^2}\right)-l}
\end{equation}
(again, $l=-\ln p_1$).
Then we obtain
\begin{eqnarray}\label{eq:803}
\frac{1}{p_2}+\frac{p_2}{2 a^2}\approx\frac{1}{p_0}-\frac{\Delta p}{p_0^2}+\frac{p_0}{2 a^2}+\frac{\Delta p}{2 a^2}=
\nonumber\\
=\frac{1}{p_0}+\frac{p_0}{2 a^2}+\Delta p\left(-\frac{1}{p_0^2}+\frac{1}{2 a^2}\right),
\end{eqnarray}
so
\begin{widetext}
\begin{eqnarray}\label{eq:804}
\eta\approx\textrm{Im} \frac{4}{\frac{1}{\Delta p}\left(1+\frac{1}{2}\frac{\Delta p}{p_0}\right)\left(\frac{1}{p_0}+\frac{p_0}{2 a^2}+\Delta p\left(-\frac{1}{p_0^2}+\frac{1}{2 a^2}\right)\right)-l}\approx\textrm{Im} \frac{4}{\frac{1}{\Delta p}\left(\frac{1}{p_0}+\frac{p_0}{2 a^2}\right)+\frac{1}{2 p_0^2}+\frac{1}{4 a^2}-\frac{1}{p_0^2}+\frac{1}{2 a^2}-l}=
\nonumber\\
=\textrm{Im} \frac{4}{\frac{1}{\Delta p}\left(\frac{1}{p_0}+\frac{p_0}{2 a^2}\right)-\frac{1}{2 p_0^2}+\frac{3}{4 a^2}-l}=\textrm{Im} \frac{4}{\frac{1}{\Delta p}Z_1+Z_2-l}=\textrm{Im} \frac{4\Delta p}{Z_1+(Z_2-l)\Delta p}=
\nonumber\\
=\textrm{Im} \frac{4\Delta p(Z_1+(Z_2-l)\Delta p^*)}{(Z_1+(Z_2-l)\Delta p)(Z_1+(Z_2-l)\Delta p^*)}=\frac{4\Delta p'' Z_1}{(Z_1+(Z_2-l)\Delta p')^2+(Z_2-l)^2\Delta p''^2}=
\nonumber\\
=\frac{4 Z_1}{(Z_2-l)^2}\frac{\Delta p''}{\left(\left(\frac{Z_1}{Z_2-l}\right)+\Delta p'\right)^2+\Delta p''^2},
\end{eqnarray}
\end{widetext}
where $Z_1$, $Z_2$, $\Delta p'$, $\Delta p''$ are real,
\begin{equation}\label{eq:805}
Z_1=\frac{1}{p_0}+\frac{p_0}{2 a^2},
\end{equation}
\begin{equation}\label{eq:806}
Z_2=-\frac{1}{2 p_0^2}+\frac{3}{4 a^2},
\end{equation}
\begin{equation}\label{eq:807}
\Delta p'+i \Delta p''=\Delta p.
\end{equation}
As $l$ is defined with an accuracy of unity, let us consider two cases:
\paragraph{\label{sec:level4221}$|Z_2-l|\sim 1$.}
In this case we obtain
\begin{equation}\label{eq:808}
\frac{3}{4 a^2}\approx l,
\end{equation}
\begin{equation}\label{eq:809}
Z_1\agt l,
\end{equation}
\begin{equation}\label{eq:810}
\eta\sim\frac{4 Z_1\Delta p''}{Z_1^2+\Delta p''^2}.
\end{equation}
If $\Delta p''\ll 1$, then $\eta\ll\frac{4}{Z_1}\alt\frac{1}{l}$. If $\Delta p''\sim 1$, then Eqs.~(\ref{eq:805},\ref{eq:808}) yield $Z_1\approx\frac{2}{3}l p_0$, so $\eta\sim\frac{1}{l}$ if $\Delta p''\sim 1$, $p_0\sim 1$, $|\Delta p'|\alt 1$.
\paragraph{\label{sec:level4222}$|Z_2-l|\gg 1$.}
As in this case
\begin{equation}\label{eq:811}
|-\frac{1}{2 p_0^2}+\frac{3}{4 a^2}-l|\gg 1,
\end{equation}
we obtain
\begin{equation}\label{eq:812}
|\frac{3}{4 a^2}-l|\gg 1.
\end{equation}
Let us consider two cases:
\subparagraph{\label{sec:level52221}$\frac{1}{a^2}\agt l$.}

In this case we obtain
\begin{equation}\label{eq:813}
Z_1\approx\frac{p_0}{2 a^2}\agt p_0 l,
\end{equation}
\begin{equation}\label{eq:814}
Z_2\approx\frac{3}{4 a^2},
\end{equation}
\begin{equation}\label{eq:815}
\left|\frac{Z_1}{Z_2-l}\right|\agt p_0,
\end{equation}
\begin{equation}\label{eq:816}
\left|\frac{Z_1}{Z_2-l}+\Delta p'\right|\agt p_0,
\end{equation}
\begin{equation}\label{eq:817}
\frac{Z_1}{(Z_2-l)^2}\sim \frac{\frac{p_0}{2 a^2}}{\left(\frac{3}{4 a^2}\right)^2}\sim p_0 a^2
\end{equation}
(we use reasoning similar to that after Eq.~(\ref{eq:785}) and assume that $|Z_2-l|\gg1$ implies $|Z_2-l|\agt l$, so $|Z_2-l|\sim Z_2$).

If $\Delta p''\ll 1$, then
\begin{equation}\label{eq:818}
\eta\ll 4 p_0 a^2\frac{1}{p_0^2}=\frac{4 a^2}{p_0}\alt\frac{4}{l}.
\end{equation}
If, on the other hand, $\Delta p''\sim 1$ , then
\begin{equation}\label{eq:818b}
\eta\sim\frac{4 a^2}{p_0}\alt\frac{4}{l},
\end{equation}
So the absorption efficiency is high if
\begin{equation}\label{eq:819}
\frac{a^2}{p_0}\sim \frac{1}{l},
\end{equation}
therefore,
\begin{equation}\label{eq:820}
a^2 l\sim 1
\end{equation}
and
\begin{equation}\label{eq:821}
p_0\sim 1.
\end{equation}
\subparagraph{\label{sec:level52222}$\frac{1}{a^2}\ll l$.}

In this case
\begin{equation}\label{eq:822}
|Z_2|\ll l
\end{equation}
and
\begin{equation}\label{eq:823}
\eta\approx\frac{4 Z_1}{l^2}\frac{\Delta p''}{\left(\left(\frac{Z_1}{Z_2-l}\right)+\Delta p'\right)^2+\Delta p''^2}.
\end{equation}
Let us determine the accuracy with which expression
\begin{equation}\label{eq:824}
\frac{Z_1}{Z_2-l}
\end{equation}
is defined:
\begin{eqnarray}\label{eq:825}
\Delta\left(\frac{Z_1}{Z_2-l}\right)\sim \frac{Z_1}{Z_2-l-1}-\frac{Z_1}{Z_2-l}\approx
\nonumber\\
\approx\frac{Z_1}{(Z_2-l)^2}\sim\frac{Z_1}{l^2}.
\end{eqnarray}
Therefore let us consider two cases:

$\left|\frac{Z_1}{Z_2-l}+\Delta p'\right|\sim\frac{Z_1}{l^2}$.

In this case
\begin{equation}\label{eq:826}
\eta\sim\frac{4 Z_1}{l^2}\frac{\Delta p''}{\left(\frac{Z_1}{l^2}\right)^2+\Delta p''^2}.
\end{equation}
In the maximum
\begin{equation}\label{eq:827}
\Delta p''\sim\frac{Z_1}{l^2},
\end{equation}
\begin{equation}\label{eq:828}
\eta\sim\frac{4 Z_1}{l^2}\frac{1}{2\frac{Z_1}{l^2}}\sim 1.
\end{equation}
There should also be
\begin{equation}\label{eq:829}
|\Delta p'|\approx\left|\frac{Z_1}{Z_2-l}\right|\alt 1.
\end{equation}
Condition
\begin{equation}\label{eq:830}
\left|\frac{Z_1}{Z_2-l}\right|\alt 1
\end{equation}
is equivalent to the following:
\begin{equation}\label{eq:831}
\frac{\frac{p_0}{2 a^2}}{l}=\frac{p_0}{2 a^2 l}\alt 1.
\end{equation}

$\left|\frac{Z_1}{Z_2-l}+\Delta p'\right|\gg\frac{Z_1}{l^2}$.

In the maximum we have
\begin{equation}\label{eq:832}
\Delta p''\approx\left|\frac{Z_1}{Z_2-l}+\Delta p'\right|\gg\frac{Z_1}{l^2},
\end{equation}
\begin{eqnarray}\label{eq:833}
\eta\approx\frac{4 Z_1}{(Z_2-l)^2}\frac{1}{2\left|\frac{Z_1}{Z_2-l}+\Delta p'\right|}=
\nonumber\\
=\frac{2}{|Z_2-l|\left|1+\Delta p'\frac{Z_2-l}{Z_1}\right|}\sim\frac{2}{l}\frac{1}{\left|1+\Delta p'\frac{Z_2-l}{Z_1}\right|}
\end{eqnarray}
In view of Eq.~(\ref{eq:832}) we may assume that
\begin{equation}\label{eq:834}
\Delta p''\agt\frac{Z_1}{l},
\end{equation}
so there should be
\begin{equation}\label{eq:834b}
\frac{Z_1}{l}\alt 1,
\end{equation}
or
\begin{equation}\label{eq:835}
\frac{p_0}{2 a^2}\alt l.
\end{equation}
In view of Eq.~(\ref{eq:833}) we have $|\Delta p'|\frac{l}{Z_1}\alt 1$, or  $|\Delta p'|\frac{2 a^2 l}{p_0}\alt 1$, otherwise $\eta\ll\frac{1}{l}$.
\subsubsection{\label{sec:level323}Parameter $p_2$ in the vicinity of a zero of $J_0(x)$}
Let us consider the case where $p_2=q_0+\Delta q$, $q_0$ being a zero of function $J_0(x)$, so $q_0$ is real and positive (negative zeros are not relevant in view of Eqs.~(\ref{eq:701}-\ref{eq:701a})),  $|\Delta q|\alt 1$, $\Delta q=\Delta q'+i \Delta q''$, where $\Delta q'$ and $\Delta q''$ are real, and $\Delta q''>0$. Using Eqs.~(\ref{eq:795}-\ref{eq:796}), we obtain
\begin{equation}\label{eq:836}
J_0(q_0)=0,
\end{equation}
\begin{equation}\label{eq:836a}
J'_0(q_0)=-J_1(q_0).
\end{equation}
Therefore,
\begin{eqnarray}\label{eq:839}
\frac{J_0(q_0+\Delta q)}{J_1(q_0+\Delta q)}\approx\frac{J_0(q_0)+\Delta q J'_0(q_0)}{J_1(q_0)}
=-\Delta q,
\end{eqnarray}
\begin{equation}\label{eq:840}
\frac{1}{p_2}+\frac{p_2}{2 a^2}\approx\frac{1}{q_0}+\frac{q_0}{2 a^2}.
\end{equation}
We leave fewer terms in the expansions than in some of the above cases, as function $\frac{J_0(x)}{J_1(x)}$, obviously, varies more slowly in the vicinity of zeros of function $J_0(x)$ than in the vicinity of zeros of function $J_1(x)$.

Let us denote $Y_1=\frac{1}{q_0}+\frac{q_0}{2 a^2}$, then
\begin{eqnarray}\label{eq:841}
\eta\approx\textrm{Im} \frac{4}{\frac{J_0(p_2)}{J_1(p_2)}\left(\frac{1}{p_2}+\frac{p_2}{2 a^2}\right)-l}\approx
\nonumber\\
\approx\textrm{Im} \frac{4}{-\Delta q\left(\frac{1}{q_0}+\frac{q_0}{2 a^2}\right)-l}=\textrm{Im} \frac{4}{-\Delta q Y_1-l}=
\nonumber\\
=\textrm{Im} \frac{4((-\Delta q'+i\Delta q'')Y_1-l)}{(-\Delta q' Y_1-l)^2+(-\Delta q'' Y_1)^2}=
\nonumber\\
=\frac{4\Delta q'' Y_1}{(\Delta q' Y_1+l)^2+\Delta q''^2 Y_1^2}.
\end{eqnarray}
There should be $Y_1\gg 1$, as otherwise $\eta\alt\frac{1}{l^2}$. Therefore,
\begin{equation}\label{eq:842}
Y_1=\frac{1}{q_0}+\frac{q_0}{2 a^2}\approx\frac{q_0}{2 a^2}\gg 1.
\end{equation}
Let us consider two cases:
\paragraph{\label{sec:level4231}$|\Delta q' \frac{q_0}{2 a^2}+l|\sim 1$.}
In the maximum
\begin{equation}\label{eq:843}
\Delta q'' \frac{q_0}{2 a^2}\sim 1,
\end{equation}
\begin{equation}\label{eq:843a}
\eta\sim 1,
\end{equation}
\begin{equation}\label{eq:844}
\left|\Delta q' \frac{q_0}{2 a^2}+l\right|\sim 1,
\end{equation}
\begin{equation}\label{eq:845}
\Delta q' \frac{q_0}{2 a^2}\approx-l.
\end{equation}
As $|\Delta q'|\alt 1$,
\begin{equation}\label{eq:846}
\frac{q_0}{2 a^2}\agt l.
\end{equation}
\paragraph{\label{sec:level3232}$|\Delta q' \frac{q_0}{2 a^2}+l|\gg 1$.}
In the vicinity of the maximum
\begin{equation}\label{eq:847}
\Delta q'' \frac{q_0}{2 a^2}\approx|\Delta q' \frac{q_0}{2 a^2}+l|\gg 1,
\end{equation}
\begin{equation}\label{eq:848}
\eta\sim\frac{2}{|\Delta q' \frac{q_0}{2 a^2}+l|},
\end{equation}
\begin{equation}\label{eq:849}
|\Delta q' \frac{q_0}{2 a^2}|\alt l,
\end{equation}
as otherwise
\begin{equation}\label{eq:850}
\eta\ll\frac{1}{l}.
\end{equation}
We may assume that
\begin{equation}\label{eq:851}
\Delta q'' \frac{q_0}{2 a^2}\agt l,
\end{equation}
as $\Delta q'' \frac{q_0}{2 a^2}\gg 1$. Therefore, $|\Delta q'|\alt\Delta q''$. On the other hand, Eq.~(\ref{eq:847}) yields
\begin{equation}\label{eq:852}
\Delta q'' \frac{q_0}{2 a^2}\sim l,
\end{equation}
so
\begin{equation}\label{eq:854}
\frac{2 a^2 l}{q_0}\alt 1.
\end{equation}
\subsection{\label{sec:level223}Domains of parameters providing high absorption efficiency (summary)}
For easy reference, let us summarize the notation:

$a$ is the radius of the cylinder (a system of units is used where the wave vector in free space $k_{0}=\frac{\omega}{c}=1$),

$p_1\approx\frac{2 a}{r_1}\ll 1$, where $r_1$ is the beam waist radius defined by $e$-fold field intensity attenuation,

$l=|\ln p_1|\ll 1$,

$p_2^2=(\varepsilon-1)a^2+p_1^2\approx (\varepsilon-1)a^2$, $p_2=p'+i p''$, $p'$ and $p''$ are real, $p''>0$,

$\varepsilon=\varepsilon'+i \varepsilon''$ is the complex permittivity of the cylinder, $\varepsilon'$ and $\varepsilon''$ are real,

$x=\frac{1}{a^2 l}$,

$y=\varepsilon'-1$,

$p_0$ is a positive zero of the Bessel function $J_1(x)$,

$\Delta p=\Delta p'+i \Delta p''=p_2-p_0$, $\Delta p'$ and $\Delta p''$ are real,

$Z_1=\frac{1}{p_0}+\frac{p_0}{2 a^2}$,

$Z_2=-\frac{1}{2 p_0^2}+\frac{3}{4 a^2}$,

$q_0$ is a positive zero of the Bessel function $J_0(x)$,

$\Delta q=\Delta q'+i \Delta q''=p_2-q_0$, $\Delta q'$ and $\Delta q''$ are real,

$Y_1=\frac{1}{q_0}+\frac{q_0}{2 a^2}$.
Let us now summarize the conditions of high absorption efficiency using a hierarchical structure:

1. $|p_2|\alt 1$

1.1.
\begin{eqnarray}\label{eq:854a}
|x-1|\sim \frac{1}{l},
\nonumber\\
\varepsilon''\sim l,
\nonumber\\
y\alt l.
\end{eqnarray}

1.2. $|x-1|\gg\frac{1}{l}$

1.2.1.

\begin{eqnarray}\label{eq:854b}
x\alt 1,
\nonumber\\
|x-1|\gg\frac{1}{l},
\nonumber\\
\varepsilon''\sim\max\left(\left|y+\frac{2 x}{x-1}\right|,\frac{2 x}{(x-1)^2 l}\right),
\nonumber\\
|y|\alt\frac{x}{(x-1)^2},
\nonumber\\
|y|\alt\frac{1}{a^2}.
\end{eqnarray}

1.2.2.
\begin{eqnarray}\label{eq:854c}
x\gg 1,
\nonumber\\
|y+2|\alt\frac{1}{x},
\nonumber\\
\varepsilon''\sim\max\left(\left|y+\frac{2 x}{x-1}\right|,\frac{2 x}{(x-1)^2 l}\right),
\nonumber\\
\frac{1}{p_1^2}\agt\frac{1}{a^4}.
\end{eqnarray}

2. $|p_2|\agt 1$

2.1. $p''\agt 1$

2.1.1.
\begin{eqnarray}\label{eq:854d}
|y+4 a^2 l^2|\alt 4 a^2 l,
\nonumber\\
8 a^2\alt\varepsilon''\alt 8 a^2 l^2,
\nonumber\\
l\agt \frac{1}{2 a^2}.
\end{eqnarray}

2.1.2.1.
\begin{eqnarray}\label{eq:854e}
|y|\alt 4 a^2 l^2,
\nonumber\\
\varepsilon''\sim 8 a^2 l^2,
\nonumber\\
l\agt \frac{1}{2 a^2}.
\end{eqnarray}

2.1.2.2.
\begin{eqnarray}\label{eq:854f}
y\sim 4 a^2 l^2,
\nonumber\\
4 l\alt\varepsilon''\alt 8 a^2 l^2.
\end{eqnarray}

2.2. $p_2=p_0+\Delta p$, $J_1(p_0)=0$, $p_0>0$, $\Delta p=\Delta p'+i \Delta p''$

2.2.1. $|Z_2-l|\sim 1$
\begin{eqnarray}\label{eq:854g}
\left|\frac{3}{4 a^2}-l\right|\sim 1,
\nonumber\\
\Delta p''\sim 1,
\nonumber\\
p_0\sim 1,
\nonumber\\
|\Delta p'|\alt 1.
\end{eqnarray}

2.2.2. $|Z_2-l|\gg 1$

2.2.2.1. $\frac{1}{a^2}\agt l$
\begin{eqnarray}\label{eq:854h}
\left|\frac{3}{4 a^2}-l\right|\gg 1,
\nonumber\\
\Delta p''\sim 1,
\nonumber\\
\Delta p'\alt 1,
\nonumber\\
a^2 l\sim 1,
\nonumber\\
p_0\sim 1.
\end{eqnarray}

2.2.2.2. $\frac{1}{a^2}\ll l$

2.2.2.2.1. $\left|\frac{Z_1}{Z_2-l}+\Delta p'\right|\sim\frac{Z_1}{l^2}$
\begin{eqnarray}\label{eq:854i}
\left|\frac{3}{4 a^2}-l\right|\gg 1,
\nonumber\\
\frac{1}{a^2}\ll l,
\nonumber\\
\left|\frac{Z_1}{Z_2-l}+\Delta p'\right|\sim\frac{Z_1}{l^2},
\nonumber\\
\Delta p''\sim\frac{Z_1}{l^2},
\nonumber\\
\frac{p_0}{2 a^2 l}\alt 1.
\end{eqnarray}

2.2.2.2.2. $\left|\frac{Z_1}{Z_2-l}+\Delta p'\right|\gg\frac{Z_1}{l^2}$
\begin{eqnarray}\label{eq:854j}
\left|\frac{3}{4 a^2}-l\right|\gg 1,
\nonumber\\
\left|\frac{Z_1}{Z_2-l}+\Delta p'\right|\gg\frac{Z_1}{l^2},
\nonumber\\
\Delta p''\sim\left|\frac{Z_1}{Z_2-l}+\Delta p'\right|,
\nonumber\\
\frac{p_0}{2 a^2}\alt l,
\nonumber\\
|\Delta p'|\frac{2 a^2 l}{p_0}\alt 1.
\end{eqnarray}

2.3. $p_2=q_0+\Delta q$, $J_0(q_0)=0$, $q_0>0$, $\Delta q=\Delta q'+i \Delta q''$

2.3.1. $\left|\Delta q'\frac{q_0}{2 a^2}+l\right|\sim 1$
\begin{eqnarray}\label{eq:854k}
\left|\Delta q'\frac{q_0}{2 a^2}+l\right|\sim 1,
\nonumber\\
\Delta q''\frac{q_0}{2 a^2}\sim 1,
\nonumber\\
\frac{q_0}{2 a^2}\agt l,
\nonumber\\
|\Delta q'|\alt 1,
\nonumber\\
\Delta q''\alt 1.
\end{eqnarray}

2.3.2. $\left|\Delta q'\frac{q_0}{2 a^2}+l\right|\gg 1$
\begin{eqnarray}\label{eq:854l}
\left|\Delta q'\frac{q_0}{2 a^2}+l\right|\gg 1,
\nonumber\\
\Delta q''\sim\left|\Delta q'+\frac{2 a^2 l}{q_0}\right|,
\nonumber\\
|\Delta q'|\frac{q_0}{2 a^2}\alt l,
\nonumber\\
\frac{q_0}{2 a^2}\agt l.
\end{eqnarray}
\maketitle

\section{\label{sec:level13}A relatively simple exact formula for power absorbed in the cylinder}
Rather surprisingly, it is possible to derive an exact formula expressing the power absorbed in the cylinder as a one-dimensional integral. It should be noted that this formula has an obvious drawback: it gives an expression for $W^a_\infty$ - power absorbed in the whole infinite cylinder ($-\infty<z<\infty$). On the other hand, as we are interested in high specific power input, it is more important for practical purposes to know the power absorbed over a finite section of the cylinder, e.g. the section where the linear absorbed power (absorbed power per unit length) is at least half as large as in the maximum ($W^a_{0.5}$). It seems, however, that this formula may be quite useful for testing of more detailed calculations. Furthermore, it is possible to get a rough estimate of ratio $\frac{W^a_{0.5}}{W^a_\infty}$. It is well-known that $z$-dependence of the magnitude of the Poynting  vector on the axis in a Gaussian beam is described by factor $\frac{1}{w_0^4+z^2}$ (as always, we use the system of units where the magnitude of the wave vector in vacuum equals unity). Thus, this factor is half as large as in the maximum at $z=\pm w_0^2$. Therefore, it is reasonable to believe that
\begin{equation}\label{eq:855}
\frac{W^a_{0.5}}{W^a_\infty}\approx\frac{\int^{w^2_0}_{-w^2_0}\frac{dz}{w_0^4+z^2}}{\int^{\infty}_{-\infty}\frac{dz}{w_0^4+z^2}}=\frac{1}{2}.
\end{equation}
Direct calculations typically yield values around $0.6$ for this ratio.

So let us derive the formula. Again, we describe the incident Gaussian beam by $z$-components of the electric and magnetic Hertz vectors:
\begin{equation}\label{eq:856}
\Pi(\rho,\varphi,z)=\int^1_0 \alpha h(\lambda)J_1(\lambda\varrho)\,d\lambda,
\end{equation}
\begin{equation}\label{eq:857}
\Pi'(\rho,\varphi,z)=\int^1_0 \beta h(\lambda)J_1(\lambda\varrho)\,d\lambda,
\end{equation}
where $\alpha=1$, $\beta=-i$,
\begin{equation}\label{eq:858}
h(\lambda)=w_0^2\exp(i(\varphi+\gamma z))\exp(-w_0^2\frac{\lambda^2}{2}),
\end{equation}
and $\gamma=\sqrt{1-\lambda^2}$ (see Eqs.(22) of Part I of this work (Ref.~\cite{Akhm10})). Then z-components of the electric and magnetic Hertz vectors  for the refracted field are:
\begin{equation}\label{eq:859}
u_2(\rho,\varphi,z)=\int^1_0 a_2(\lambda) h(\lambda)J_1(\lambda_2\varrho)\,d\lambda,
\end{equation}
\begin{equation}\label{eq:860}
v_2(\rho,\varphi,z)=\int^1_0 b_2(\lambda) h(\lambda)J_1(\lambda_2\varrho)\,d\lambda,
\end{equation}
where $a_2(\lambda)$ and $b_2(\lambda)$ are given by Eqs.(99,101) of (Ref.~\cite{Akhm10}) for $n=1$, $\varepsilon_1=\mu_1=\mu_2=1$, $\varepsilon_2=\varepsilon$, $p_1=\lambda a$, and $\lambda_2=\varepsilon-\gamma^2$, $p_2=\lambda_2 a$. These Hertz vectors yield the following refracted electric and magnetic fields on the surface of the cylinder ($\rho=a$):
\begin{widetext}
\begin{eqnarray}\label{eq:861}
\bm{H}(a,\varphi,z)=\int^1_0 a_2(\lambda) h(\lambda)\left\{\frac{1}{a}J_1(p_2),i\lambda_2 J'_1(p_2),0\right\}\,d\lambda,
\end{eqnarray}
\begin{eqnarray}\label{eq:862}
\bm{E}(a,\varphi,z)=\int^1_0 a_2(\lambda) h(\lambda)\left\{\frac{i \gamma\lambda_2}{\varepsilon}J'_1(p_2),-\frac{\gamma}{\varepsilon a} J_1(p_2),\frac{\lambda_2^2}{\varepsilon}J_1(p_2)\right\}\,d\lambda,
\end{eqnarray}
\begin{eqnarray}\label{eq:863}
\bm{H'}(a,\varphi,z)=\int^1_0 b_2(\lambda) h(\lambda)\left\{i \gamma\lambda_2 J'_1(p_2),-\frac{\gamma}{a} J_1(p_2),\lambda_2^2 J_1(p_2)\right\}\,d\lambda,
\end{eqnarray}
\begin{eqnarray}\label{eq:864}
\bm{E'}(a,\varphi,z)=\int^1_0 b_2(\lambda) h(\lambda)\left\{-\frac{1}{a}J_1(p_2),-i \lambda_2 J'_1(p_2),0\right\}\,d\lambda
\end{eqnarray}
\end{widetext}
(cf. Eqs.(78-79) of (Ref.~\cite{Akhm10}). The total refracted electric and magnetic fields are $\bm{E}^{r\!fr}=\bm{E}+\bm{E'}$ and $\bm{H}^{r\!fr}=\bm{H}+\bm{H'}$, correspondingly. The averaged $\rho$-component of the Poynting vector at the surface of the cylinder equals
\begin{eqnarray}\label{eq:865}
\frac{1}{2}\frac{c}{4 \pi}\textrm{Re}\left[\bm{E}^{r\!fr}(a,\varphi,z)\times\bm{H}^{r\!fr*}(a,\varphi,z)\right]_\rho=
\nonumber\\
=\frac{1}{2}\frac{c}{4 \pi}\textrm{Re}\left(E^{r\!fr}_\varphi H^{r\!fr*}_z-E^{r\!fr}_z H^{r\!fr*}_\varphi\right).
\end{eqnarray}
The total power absorbed in the cylinder equals
\begin{eqnarray}\label{eq:866}
W^a_\infty=-a\int_0^{2 \pi}d\varphi\int_{-\infty}^{\infty}dz
\nonumber\\
\times\frac{1}{2}\frac{c}{4 \pi}\textrm{Re}\left(E^{r\!fr}_\varphi H^{r\!fr*}_z-E^{r\!fr}_z H^{r\!fr*}_\varphi\right).
\end{eqnarray}
We obtain:
\begin{widetext}
\begin{eqnarray}\label{eq:867}
E^{r\!fr}_\varphi=
\int^1_0 \left(-a_2(\lambda)\frac{\gamma}{\varepsilon a}J_1(p_2)-i b_2(\lambda)\lambda_2 J'_1(p_2)\right)  h(\lambda)\,d\lambda,
\end{eqnarray}
\begin{eqnarray}\label{eq:868}
E^{r\!fr}_z=
\int^1_0 a_2(\lambda)\frac{\lambda_2^2}{\varepsilon}J_1(p_2)  h(\lambda)\,d\lambda,
\end{eqnarray}
\begin{eqnarray}\label{eq:869}
H^{r\!fr*}_\varphi=
\int^1_0 \left(-i a_2^*(\lambda)\lambda^*_2J'^*_1(p_2)-b_2^*(\lambda)\frac{\gamma}{a}J^*_1(p_2)\right)  h^*(\lambda)\,d\lambda,
\end{eqnarray}
\begin{eqnarray}\label{eq:870}
H^{r\!fr*}_z=
\int^1_0 b_2^*(\lambda)\lambda^{*2}_2J^*_1(p_2)  h^*(\lambda)\,d\lambda.
\end{eqnarray}
\end{widetext}
We also have
\begin{eqnarray}\label{eq:871}
\int^1_0\ldots d\lambda=-\int^0_1\ldots\frac{\gamma\,d\gamma}{\lambda}=\int^1_0\ldots\frac{\gamma\,d\gamma}{\lambda}.
\end{eqnarray}
Integration with respect to $\varphi$ in Eq.~(\ref{eq:866}) yields factor $2 \pi$. The resulting three-dimensional integral may be reduced to a one-dimensional one using the following reasoning:
\begin{widetext}
\begin{eqnarray}\label{eq:872}
\int_{-\infty}^\infty dz\int_0^1 g_1(\gamma)\exp(i\gamma z)\,d\gamma\int_0^1 g_2(\gamma')\exp(-i\gamma' z)\,d\gamma'=
\int_0^1\int_0^1d\gamma d\gamma' g_1(\gamma)g_2(\gamma')\int_{-\infty}^\infty dz \exp(i(\gamma-\gamma')z)=
\nonumber\\
=\int_0^1\int_0^1d\gamma d\gamma' g_1(\gamma)g_2(\gamma')\cdot 2\pi\delta(\gamma-\gamma')=2 \pi\int_0^1d\gamma g_1(\gamma)g_2(\gamma).
\end{eqnarray}
Therefore,
\begin{eqnarray}\label{eq:873}
W^a_\infty=-a\cdot 2\pi\cdot\frac{1}{2}\frac{c}{4\pi}\cdot 2\pi\int_0^1 \left(w_0^2\exp\left(-w_0^2\frac{\lambda^2}{2}\right)\frac{\gamma}{\lambda}\right)^2
\chi(\lambda)\,d\gamma=-a\frac{\pi c}{2}w_0^4\int_0^1 \exp(-w_0^2\lambda^2)\frac{\gamma}{\lambda}\chi(\lambda)\,d\lambda,
\end{eqnarray}
where
\begin{eqnarray}\label{eq:874}
\chi(\lambda)=\textrm{Re}\left(\left(-a_2(\lambda)\frac{\gamma}{\varepsilon a}J_1(p_2)-i b_2(\lambda)\lambda_2 J'_1(p_2)\right)b_2^{*}(\lambda)\lambda_2^{*2}J_1^{*}(p_2)\right)-
\nonumber\\
-\textrm{Re}\left(a_2(\lambda)\frac{\lambda_2^2}{\varepsilon}J_1(p_2)\left(-i a_2^{*}(\lambda)\lambda_2^{*} J'^{*}_1(p_2)-b_2^{*}(\lambda)\frac{\gamma}{a}J_1^{*}(p_2)\right)\right)=
\nonumber\\
=J_1(p_2)J^*_1(p_2)\textrm{Re}\left(-a_2(\lambda)b^*_2(\lambda)\frac{\gamma}{\varepsilon a}\lambda^{*2}_2+a_2(\lambda)b^*_2(\lambda)\frac{\lambda_2^2}{\varepsilon}\frac{\gamma}{a}\right)+
\nonumber\\
+\textrm{Re}\left(J'_1(p_2)J^*_1(p_2)(-i b_2(\lambda)b^*_2(\lambda)\lambda_2\lambda^{*2}_2)+J_1(p_2)J'^*_1(p_2)\left(i a_2(\lambda)a^*_2(\lambda)\frac{\lambda_2^2}{\varepsilon}\lambda^{*}_2\right)\right)=
\nonumber\\
=J_1(p_2)J^*_1(p_2)\textrm{Re}\left(a_2(\lambda)b^*_2(\lambda)\frac{\gamma}{\varepsilon a}(-\lambda^{*2}_2+\lambda^2_2)+\frac{J'_1(p_2)}{J_1(p_2)}\left(-i b_2(\lambda)b^*_2(\lambda)\lambda_2\lambda^{*2}_2-ia^*_2(\lambda)a_2(\lambda)\frac{\lambda^{*2}}{\varepsilon^*}\lambda_2\right)\right).
\end{eqnarray}
\end{widetext}
As $\varepsilon-\lambda^2_2=\gamma^2$, we have $\lambda^2_2-\lambda^{*2}_2=\varepsilon-\varepsilon^*$. The following property of the Bessel functions
\begin{equation}\label{eq:875}
J_0(x)=\frac{J_1(x)}{x}+J'_1(x)
\end{equation}
yields
\begin{equation}\label{eq:876}
x\frac{J'_1(x)}{J_1(x)}=x\frac{J_0(x)}{J_1(x)}-1,
\end{equation}
therefore,
\begin{widetext}
\begin{eqnarray}\label{eq:877}
\chi(\lambda)=\frac{1}{a}|J_1(p_2)|^2\textrm{Re}\left(a_2(\lambda)b^*_2(\lambda)\gamma\left(1-\frac{\varepsilon^*}{\varepsilon}\right)-i\left(p_2\frac{J_0(p_2)}{J_1(p_2)}-1\right)
(\varepsilon^*-\gamma^2)\left(|b_2(\lambda)|^2+\frac{1}{\varepsilon^*}|a_2(\lambda)|^2\right)\right).
\end{eqnarray}
\end{widetext}
There are some reasons to believe that a similar procedure can yield an exact formula for the power in the incident Gaussian beam, but in this work the power in the beam is evaluated using an asymptotic formula for $w_0\gg 1$. As Eqs.~(\ref{eq:856}-\ref{eq:858}) define a solution of the free Maxwell equation that is accurately approximated by a Gaussian beam for $w_0\gg 1$, we shall use the expressions for the electric and magnetic fields of the Gaussian beam at $z=0$ (Eqs.(7-8) of (Ref.~\cite{Akhm10}). The averaged $\rho$-component of the Poynting vector at the point$\{x,y,z\}$ ($z=0$) equals
\begin{eqnarray}\label{eq:878}
\frac{1}{2}\frac{c}{4 \pi}\textrm{Re}\left[\bm{E}^{inc}\{x,y,0\}\times\bm{H}^{inc*}\{x,y,0\}\right]_z=
\nonumber\\
=\frac{1}{2}\frac{c}{4 \pi}\textrm{Re}\left(E^{inc}_x H^{inc*}_y-E^{inc}_y H^{inc*}_x\right)=
\nonumber\\
=\frac{1}{2}\frac{c}{4 \pi}\textrm{Re}(i\cdot(-i)+1)\exp\left(-\frac{x^2+y^2}{w_0^2}\right)=
\nonumber\\
=\frac{c}{4 \pi}\exp\left(-\frac{x^2+y^2}{w_0^2}\right).
\end{eqnarray}
Therefore, the power in the incident beam equals
\begin{eqnarray}\label{eq:879}
\frac{c}{4 \pi}\int^{\infty}_{-\infty}\int^{\infty}_{-\infty}\exp\left(-\frac{x^2+y^2}{w_0^2}\right)dx dy=
\nonumber\\
=\frac{c}{4 \pi}\int^{2\pi}_{0}d\varphi\int^{\infty}_{0}r dr\exp\left(-\frac{r^2}{w_0^2}\right)=
\nonumber\\
=\frac{c}{4 \pi}2 \pi\int^{\infty}_{0}\frac{w^2_0 du}{2}\exp(-u)=\frac{c w_0^2}{4}
\end{eqnarray}
(in Eqs.(7-8) of (Ref.~\cite{Akhm10} we replace $\delta$ with $w_0$).

Thus, Eqs.~(\ref{eq:873},\ref{eq:877},\ref{eq:879}) yield an expression for heating efficiency:
\begin{eqnarray}\label{eq:880}
\eta_{n\!ew}=-2\pi w_0^2\int_0^1 \exp(-w_0^2\lambda^2)\frac{\gamma}{\lambda}\psi(\lambda)\,d\lambda,
\end{eqnarray}
where
\begin{widetext}
\begin{eqnarray}\label{eq:881}
\psi(\lambda)=|J_1(p_2)|^2\textrm{Re}\left(a_2(\lambda)b^*_2(\lambda)\gamma\left(1-\frac{\varepsilon^*}{\varepsilon}\right)-i\left(p_2\frac{J_0(p_2)}{J_1(p_2)}-1\right)
(\varepsilon^*-\gamma^2)\left(|b_2(\lambda)|^2+\frac{1}{\varepsilon^*}|a_2(\lambda)|^2\right)\right).
\end{eqnarray}
\end{widetext}
Let us find a relation between this expression and Eq.~(\ref{eq:700}) for $w_0\gg 1$. The only area that contributes to the integral of Eq.~(\ref{eq:880}) is that where $w_0 \lambda \alt 1$, so in this case we may assume that $\lambda\ll 1$ and, therefore, $\gamma\approx 1$. If we also adopt the assumptions of Eqs.(107-110,132) of (Ref.~\cite{Akhm10}, we may use the following approximations (see Eq.(116) of (Ref.~\cite{Akhm10})):
\begin{eqnarray}\label{eq:882}
a_2(\lambda)\approx\frac{i}{p_2^2 J_1(p_2)p_1 D}(\alpha+i\beta)i\varepsilon=\frac{2i}{p_2^2 J_1(p_2)p_1 D}i\varepsilon,
\nonumber\\
b_2(\lambda)\approx\frac{i}{p_2^2 J_1(p_2)p_1 D}(\alpha+i\beta)=\frac{2i}{p_2^2 J_1(p_2)p_1 D}.\quad
\end{eqnarray}
Therefore,
\begin{widetext}
\begin{eqnarray}\label{eq:883}
\psi(\lambda)\approx |J_1(p_2)|^2 |b_2(\lambda)|^2\textrm{Re}\left(i\varepsilon\left(1-\frac{\varepsilon^*}{\varepsilon}\right)-i\left(p_2\frac{J_0(p_2)}{J_1(p_2)}-1\right)
(\varepsilon^*-1)(1+\varepsilon)\right)\approx
\nonumber\\
\approx\frac{4}{|p_2|^4p_1^2|D|^2}\textrm{Re}\left(i\varepsilon-i\varepsilon^*-ip_2\frac{J_0(p_2)}{J_1(p_2)}
(\varepsilon^*-\varepsilon+\varepsilon\varepsilon^*-1)+i\varepsilon^*-i\varepsilon+i\varepsilon\varepsilon^*-i\right)=
\nonumber\\
=\frac{4}{|p_2|^4p_1^2|D|^2}\textrm{Im}\left(p_2\frac{J_0(p_2)}{J_1(p_2)}(\varepsilon^*-\varepsilon+\varepsilon\varepsilon^*-1)\right).
\end{eqnarray}
\end{widetext}
Let us prove that
\begin{eqnarray}\label{eq:884}
\frac{a^2}{2|p_2|^4}\textrm{Im}\left(p_2\frac{J_0(p_2)}{J_1(p_2)}(\varepsilon^*-\varepsilon+\varepsilon\varepsilon^*-1)\right)\approx
\nonumber\\
\approx\textrm{Im}\left(\frac{1}{p_2^2}+\frac{\varepsilon+1}{2}\frac{J'_1(p_2)}{p_2J_1(p_2)}\right).
\end{eqnarray}
In fact, as
\begin{equation}\label{eq:885}
p_2^2\approx(\varepsilon-1)a^2
\end{equation}
(see Eq.~(\ref{eq:701}) of this work and Eq.(110) of (Ref.~\cite{Akhm10}), we obtain that the left-hand side of Eq.~(\ref{eq:884}) equals:
\begin{eqnarray}\label{eq:885b}
\frac{a^2}{2}\textrm{Im}\left(\frac{1}{p_2^{*2}}\frac{J_0(p_2)}{p_2 J_1(p_2)}(\varepsilon^*-1)(1+\varepsilon)\right)\approx
\nonumber\\
\approx\frac{a^2}{2}\textrm{Im}\left(\frac{1}{(\varepsilon^*-1)a^2}\frac{J_0(p_2)}{p_2 J_1(p_2)}(\varepsilon^*-1)(1+\varepsilon)\right)=
\nonumber\\
=\textrm{Im}\left(\frac{1+\varepsilon}{2}\frac{J_0(p_2)}{p_2 J_1(p_2)}\right)=
\nonumber\\
=\textrm{Im}\left(\frac{1+\varepsilon}{2}\left(\frac{1}{p_2^2}+\frac{J'_1(p_2)}{p_2 J_1(p_2)}\right)\right)=
\nonumber\\
=\textrm{Im}\left(\left(\frac{\varepsilon-1}{2}+1\right)\frac{1}{p_2^2}+\frac{1+\varepsilon}{2}\frac{J'_1(p_2)}{p_2 J_1(p_2)}\right)\approx
\nonumber\\
\approx\textrm{Im}\left(\frac{\varepsilon-1}{2}\frac{1}{(\varepsilon-1)a^2}+\frac{1}{p_2^2}+\frac{1+\varepsilon}{2}\frac{J'_1(p_2)}{p_2 J_1(p_2)}\right)=
\nonumber\\
=\textrm{Im}\left(\frac{1}{p_2^2}+\frac{1+\varepsilon}{2}\frac{J'_1(p_2)}{p_2 J_1(p_2)}\right).\quad
\end{eqnarray}
This proves Eq.~(\ref{eq:884}). Thus, Eq.~(\ref{eq:883}) yields
\begin{eqnarray}\label{eq:886}
\psi(\lambda)\approx\frac{4}{p_1^2|D|^2}\frac{2}{a^2}\textrm{Im}\left(\frac{1}{p_2^2}+\frac{1+\varepsilon}{2}\frac{J'_1(p_2)}{p_2 J_1(p_2)}\right).
\end{eqnarray}
If we denote
\begin{eqnarray}\label{eq:887}
\eta(\lambda)=-\frac{16}{p_1^4|D|^2}\textrm{Im}\left(\frac{1}{p_2^2}+\frac{1+\varepsilon}{2}\frac{J'_1(p_2)}{p_2 J_1(p_2)}\right),
\end{eqnarray}
then Eqs.(129,143) of (Ref.~\cite{Akhm10}) yield
\begin{eqnarray}\label{eq:888}
\eta(\lambda)\approx\textrm{Im} \frac{4}{p_2\frac{J_0(p_2)}{J_1(p_2)}\left(\frac{1}{p_2^2}+\frac{1}{2 a^2}\right)+\ln p_1},
\end{eqnarray}
therefore,
\begin{eqnarray}\label{eq:889}
\psi(\lambda)\approx\eta(\lambda)\frac{8 p_1^2}{(-16)a^2}\approx-\frac{1}{2}\eta(\lambda)\lambda^2,
\end{eqnarray}
and
\begin{eqnarray}\label{eq:890}
\eta_{n\!ew}\approx\pi w_0^2\int_0^1 \exp(-w_0^2\lambda^2)\gamma\lambda\eta(\lambda)\,d\lambda\approx
\nonumber\\
\approx\pi w_0^2\int_0^\infty \exp(-w_0^2\lambda^2)\eta(\lambda)\lambda\,d\lambda=\frac{\pi}{2}\bar{\eta},
\end{eqnarray}
where $\bar{\eta}$ is a weighted average of $\eta(\lambda)$:
\begin{eqnarray}\label{eq:891}
\bar{\eta}=\frac{\int_0^\infty\eta(\lambda) \exp(-w_0^2\lambda^2)\lambda\,d\lambda}{\int_0^\infty \exp(-w_0^2\lambda^2)\lambda\,d\lambda}.
\end{eqnarray}
\section{\label{sec:level15}Efficient heating in the transverse geometry (more general conditions)}
Formulae for heating efficiency in the transverse geometry were derived in the first part of this work (Ref.~\cite{Akhm10}, Eqs.(156,158)). The formulae were derived in the assumption that $\varepsilon''\gg|\varepsilon'|$. Let us remove this condition now and obtain somewhat more general formulae. Let us rewrite the formula for heating efficiency of (Ref.~\cite{Akhm10}, Eqs.(151)):
\begin{equation}\label{eq:930}
\eta=\frac{|b^2|R}{-\frac{4 i}{\pi a}},
\end{equation}
where
\begin{equation}\label{eq:931}
b=-\frac{4 i}{\pi a Q},
\end{equation}
\begin{equation}\label{eq:932}
Q=\sqrt{\varepsilon}J^{'}_0(\sqrt{\varepsilon} a)H_0^{(1)}(a)-J_0(\sqrt{\varepsilon}a)H_0^{(1)'}(a),
\end{equation}
\begin{equation}\label{eq:933}
R=J^*_0(\sqrt{\varepsilon}a)\sqrt{\varepsilon}J^{'}_0(\sqrt{\varepsilon}a)
-J_0(\sqrt{\varepsilon}a)\sqrt{\varepsilon}^*J_0^{'*}(\sqrt{\varepsilon}a).
\end{equation}
We still have
\begin{equation}\label{eq:934}
R\approx-i\varepsilon'' a
\end{equation}
(Ref.~\cite{Akhm10}, Eqs.(155)), but we need more precise approximations of the cylindrical functions of small argument to calculate $Q$:
\begin{eqnarray}\label{eq:935}
J_0(z)\approx 1-\frac{z^2}{4},
\nonumber\\
J^{'}_0(z)\approx -\frac{z}{2},
\nonumber\\
H^{1}_0(z)\approx 1+\frac{2 i}{\pi}\left(\ln\left(\frac{z}{2}\right)+C\right),
\nonumber\\
H_0^{(1)'}(z)\approx -\frac{z}{2}+\frac{2 i}{\pi z}+\frac{i z}{\pi}\left(\frac{1}{2}-\ln\left(\frac{z}{2}\right)-C\right),
\end{eqnarray}
where $C=0.5772\ldots$ is the Euler constant. We leave more terms in the expansion of the first derivative of the Hankel function as it's magnitude is large for small argument. Then, neglecting some terms of higher order in $a$ and $p=\sqrt{\varepsilon} a$, we obtain:
\begin{eqnarray}\label{eq:936}
Q\approx \sqrt{\varepsilon}\left(-\frac{\sqrt{\varepsilon} a}{2}\right)\left(1+\frac{2 i}{\pi}\left(\ln\left(\frac{a}{2}\right)+C\right)\right)-
\nonumber\\
-\left(1-\frac{\varepsilon a^2}{4}\right)\left(-\frac{a}{2}+\frac{2 i}{\pi a}+\frac{i a}{\pi}L\right)\approx
\nonumber\\
\approx -\frac{\varepsilon a}{2}\left(1+\frac{2 i}{\pi}\left(\ln\left(\frac{a}{2}\right)+C\right)\right)+\frac{a}{2}-\frac{2 i}{\pi a}-
\nonumber\\
-\frac{i a}{\pi}L+\frac{i \varepsilon a}{2 \pi}=-\frac{\varepsilon a}{2}\left(1+\frac{2 i}{\pi}\left(-L\right)\right)+
\nonumber\\
+\frac{a}{2}-\frac{2 i}{\pi a}-\frac{i a}{\pi}L=-\frac{\varepsilon a}{2}+\frac{i \varepsilon a}{\pi}L+\frac{a}{2}-
\nonumber\\
-\frac{2 i}{\pi a}-\frac{i a}{\pi}L=-\frac{a}{2}(\varepsilon-1)+\frac{i a}{\pi}L(\varepsilon-1)-\frac{2 i}{\pi a}=
\nonumber\\
=-(\varepsilon-1)a\left(\frac{1}{2}-i\frac{L}{\pi}\right)-\frac{2 i}{\pi a},
\end{eqnarray}
where
\begin{equation}\label{eq:937}
L=\frac{1}{2}-\ln\left(\frac{a}{2}\right)-C.
\end{equation}
Thus,
\begin{eqnarray}\label{eq:938}
b=-\frac{4 i}{\pi a Q}\approx\frac{-4 i}{\pi a\left(-(\varepsilon-1)a\left(\frac{1}{2}-i\frac{L}{\pi}\right)-\frac{2 i}{\pi a}\right)}=
\nonumber\\
=\frac{4}{\pi i\left(-(\varepsilon-1)a^2\left(\frac{1}{2}-i\frac{L}{\pi}\right)-\frac{2 i}{\pi}\right)}=
\nonumber\\
=\frac{4}{2-i(\varepsilon-1)a^2\left(\frac{\pi}{2}-i L\right)}=
\nonumber\\
=\frac{4}{2-(\varepsilon-1)a^2\left(L+i\frac{\pi}{2}\right)}=
\nonumber\\
=\frac{4}{2-(x_1+i x_2)\left(L+i\frac{\pi}{2}\right)}=
\nonumber\\
=\frac{4}{2-L x_1-i L x_2-i\frac{\pi}{2}x_1+\frac{\pi}{2}x_2}=
\nonumber\\
=\frac{4}{\left(2-L x_1+\frac{\pi}{2}x_2\right)-i\left(L x_2+\frac{\pi}{2}x_1\right)},
\end{eqnarray}
where $x_1$, $x_2$ are real and
\begin{equation}\label{eq:939}
x_1+i x_2=(\varepsilon-1)a^2.
\end{equation}
Then
\begin{equation}\label{eq:940}
|b^2|\approx\frac{16}{\left(2-L x_1+\frac{\pi}{2}x_2\right)^2+\left(L x_2+\frac{\pi}{2}x_1\right)^2},
\end{equation}
\begin{eqnarray}\label{eq:941}
\eta\approx\frac{16}{\left(2-L x_1+\frac{\pi}{2}x_2\right)^2+\left(L x_2+\frac{\pi}{2}x_1\right)^2}
\nonumber\\
\times\frac{\left(-i\frac{x_2}{a}\right)}{\left(-\frac{4 i}{\pi a}\right)}=
\nonumber\\
=\frac{4\pi}{\frac{1}{x_2}\left(\left(2-L x_1+\frac{\pi}{2}x_2\right)^2+\left(L x_2+\frac{\pi}{2}x_1\right)^2\right)}.
\end{eqnarray}
Let us denote the denominator of the latter expression by $A$. Then
\begin{eqnarray}\label{eq:942}
A=\frac{1}{x_2}\left(2-L x_1\right)^2+\pi (2-L x_1)+\frac{\pi^2}{4}x_2+
\nonumber\\
+L^2x_2+\pi L x_1+\frac{1}{x_2}\frac{\pi^2}{4}x_1^2=
\nonumber\\
=\frac{1}{x_2}\left(\left(2-L x_1\right)^2+\frac{\pi^2}{4}x_1^2\right)+
\nonumber\\
+2\pi+\left(L^2+\frac{\pi^2}{4}\right)x_2.
\end{eqnarray}
Let us find $x_1$ and $x_2>0$ that yield the minimum of $A$ (and, therefore, the maximum of $\eta$). Differentiating $A$ with respect to $x_1$, we obtain:
\begin{eqnarray}\label{eq:943}
\frac{1}{x_2}\left(2(2-L x_1)(-L)+\frac{\pi^2}{2}x_1\right)=0,
\end{eqnarray}
so
\begin{equation}\label{eq:944}
-4 L+2 L^2 x_1+\frac{\pi^2}{2}x_1=0,
\end{equation}
and
\begin{equation}\label{eq:945}
x_1=\frac{2 L}{L^2+\frac{\pi^2}{4}}\approx\frac{2}{L}.
\end{equation}
Differentiating $A$ with respect to $x_2$, we obtain:
\begin{equation}\label{eq:946}
-\frac{1}{x_2^2}\left(\left(2-L x_1\right)^2+\frac{\pi^2}{4}x_1^2\right)+L^2+\frac{\pi^2}{4}=0,
\end{equation}
therefore
\begin{eqnarray}\label{eq:947}
x_2^2\left(L^2+\frac{\pi^2}{4}\right)=4-4 L x_1+x_1^2\left(L^2+\frac{\pi^2}{4}\right)=
\nonumber\\
=4-4 L x_1+2 L x_1=4-2 L x_1=
\nonumber\\
=4-\frac{4 L^2}{L^2+\frac{\pi^2}{4}}=\frac{\pi^2}{L^2+\frac{\pi^2}{4}},
\end{eqnarray}
so
\begin{equation}\label{eq:948}
x_2=\frac{\pi}{L^2+\frac{\pi^2}{4}}\approx\frac{\pi}{L^2}.
\end{equation}
It is easy to see that in the maximum $\eta=1$. Reflectionless absorption of cylindrical electromagnetic waves incident on uniform cylinders with small electrical dimensions was studied in \cite{Zharov1}, but the relevant case was not mentioned there, perhaps because that article used the approximation of Hankel functions for small argument that was not applicable in the case of zero index.

Let us rewrite Eq.~(\ref{eq:942}) as follows:
\begin{widetext}
\begin{eqnarray}\label{eq:949}
A=\frac{1}{x_2}\left(4-4 L x_1+L^2 x_1^2+\frac{\pi^2}{4}x_1^2\right)+2\pi+\left(L^2+\frac{\pi^2}{4}\right)x_2=\frac{1}{x_2}\left(L^2+\frac{\pi^2}{4}\right)\left(\frac{4-4 L x_1}{L^2+\frac{\pi^2}{4}}+x_1^2\right)+
\nonumber\\
+2\pi+\left(L^2+\frac{\pi^2}{4}\right)x_2=\left(L^2+\frac{\pi^2}{4}\right)\left(\frac{1}{x_2}\left(\left(x_1-\frac{2 L}{L^2+\frac{\pi^2}{4}}\right)^2+\frac{\pi^2}{\left(L^2+\frac{\pi^2}{4}\right)^2}\right)+x_2\right)+
2\pi.
\end{eqnarray}
\end{widetext}
For $\eta\agt\frac{1}{L}$, we should have
\begin{eqnarray}\label{eq:950}
\frac{1}{x_2}\left(\left(x_1-\frac{2 L}{L^2+\frac{\pi^2}{4}}\right)^2+\frac{\pi^2}{\left(L^2+\frac{\pi^2}{4}\right)^2}\right)+x_2\alt
\nonumber\\
\alt\frac{4 \pi}{L}
\end{eqnarray}
or
\begin{widetext}
\begin{eqnarray}\label{eq:951}
\max\left\{\frac{1}{x_2}\left(x_1-\frac{2 L}{L^2+\frac{\pi^2}{4}}\right)^2,\frac{1}{x_2}\frac{\pi^2}{\left(L^2+\frac{\pi^2}{4}\right)^2},x_2\right\}
\alt
\frac{4 \pi}{L},
\end{eqnarray}
\end{widetext}
or simultaneously
\begin{eqnarray}\label{eq:952}
x_2\alt\frac{4 \pi}{L},
\nonumber\\
\frac{1}{x_2}\frac{\pi^2}{\left(L^2+\frac{\pi^2}{4}\right)^2}\alt\frac{4 \pi}{L},
\nonumber\\
\frac{1}{x_2}\left(x_1-\frac{2 L}{L^2+\frac{\pi^2}{4}}\right)^2\alt\frac{4 \pi}{L}.
\end{eqnarray}
Finally, after some transformations, we obtain the following conditions of efficient heating in the transverse geometry (these conditions are sufficient, not necessary, and describe a domain of parameters where the diameter of the cylinder is much smaller than the wavelength):
\begin{eqnarray}\label{eq:953}
x_2\alt\frac{4 \pi}{L},
\nonumber\\
x_2\agt\frac{1}{L^3,}
\nonumber\\
\left(x_1-\frac{2 L}{L^2+\frac{\pi^2}{4}}\right)^2\alt\frac{4 \pi}{L}x_2,
\end{eqnarray}\maketitle
where $x_1$, $x_2$ are real and
\begin{equation}\label{eq:954}
x_1+i x_2=(\varepsilon-1)a^2,
\end{equation}
$a\ll 1$ is the radius of the cylinder (we consider a monochromatic beam with frequency $\omega$ and use such a system of units that the wave vector in free space $k_{0}=\frac{\omega}{c}=1$); the fields in the incident beam are
described by (Ref.~\cite{Akhm10}, Eqs.(144)), $\varepsilon$ is the complex permittivity of the cylinder,
\begin{equation}\label{eq:955}
L=\frac{1}{2}-\ln\left(\frac{a}{2}\right)-C,
\end{equation}
$C=0.5772\ldots$ is the Euler constant.
\section{\label{sec:level16}Efficient laser heating of nanotubes}
Let us discuss laser heating in the transverse geometry of carbon nanotubes (e.g. multi-walled nanotubes) as the extreme case of ultrathin targets. The resulting hot dense plasma may become a lasing medium for generation of coherent extreme ultraviolet (EUV) and soft X-ray radiation. In our example, lasing may occur as a result of recombination in H-like carbon rapidly cooled by hydrodynamic expansion (Ref.~\cite{Attw1}). Laser irradiation of multiple nanotubes was previously used for X-ray generation (Ref.~\cite{Nishi2}). Let us show, however, that efficient laser heating of a single nanotube is also possible.

Let us assume that the wavelength of the pumping radiation is 800 nm  (Ti:Sapphire laser) and the pulse length is about 30 femtoseconds ($3\cdot 10^{-14}$ s), the diameter of the nanotube is 15 nm and its density is 1.3 g/cm$^2$. Nanotubes can be several centimeters long (Refs.~\cite{Zhu1},~\cite{By1}), but, due to the cylindrical symmetry of the problem, the following calculations are performed per one centimeter of the length of the nanotube.

Nanotubes may have very complex electric properties, but their atoms will be ionized rapidly in a strong electromagnetic field, so we assume that they have electric properties of plasma cylinders with complex permittivity
\begin{equation}\label{eq:956}
\varepsilon=1-\frac{\omega_p^2}{\omega(\omega+i\nu_{ei})},
\end{equation}
where
\begin{equation}\label{eq:957}
\omega_p=\left(\frac{4 \pi n_e e^2}{m_e}\right)^\frac{1}{2}
\end{equation}
is the plasma frequency, $n_e$ is the electron density (it is assumed that carbon atoms are totally ionized),
\begin{equation}\label{eq:958}
\nu_{ei}\simeq\frac{Z\omega_p}{11 N_D}\ln\frac{9 N_D}{2 Z}
\end{equation}
is the electron-ion collision frequency (Ref.~\cite{Attw1}),
\begin{equation}\label{eq:959}
N_D=\frac{4 \pi}{3}\lambda_D^3 n_e
\end{equation}
is the Debye number,
\begin{equation}\label{eq:960}
\lambda_D=\left(\frac{T_e}{4 \pi n_e e^2}\right)^\frac{1}{2}
\end{equation}
is the Debye radius, $Z=6$ is the atomic number of carbon, $T_e$ is the plasma electron temperature. If $T_e$=1 keV, the ion sound velocity is of the order of $10^7$ cm/s, so the radius of the plasma cylinder should not change dramatically over the laser pulse length.

The above simple model of plasma electromagnetic properties does not take into account nonlocality of the complex permittivity. This is partially justified by the fact that the only nonvanishing component of the electric field is directed along the axis of the cylinder, so the normal component of the electric field on the surface of the cylinder vanishes. Therefore, vacuum heating (Ref.~\cite{Brun1}) and resonance absorption (Ref.~\cite{Kru1}) should not play a major role.

The energy required for total ionizing and isochoric heating to $T_e=1$ keV equals $\frac{3}{2}(1+Z)T_e +U_{ti}$ per atom (although the ion temperature may be lower than the electron temperature) , where $U_{ti}\approx 1$ keV (Ref.~\cite{Hud1}) is the total ionization potential of a carbon atom. Thus, the energy required to heat a 1 cm long carbon nanotube with a 15 nm diameter to this temperature is about $0.2$ mJ.

The heating efficiency at $T_e=1$ keV is about 16\%. Thus, the required energy in the laser pulse may be estimated as 1.3 mJ, and the laser power is about 45 GW. The power of modern Ti:Sapphire pulsed lasers may be as high as 100 TW.
It should be noted that the heating efficiency is higher at lower temperatures. For example, the efficiency is about 40\% at $T_e=200$ eV. The efficiency is about 1\% at $T_e=10$ keV. It is possible to achieve higher efficiency at this higher temperature by varying the wavelength of the heating radiation and the nanotube diameter. It is also possible to use hydrodynamic expansion of the plasma.
\section{\label{sec:level17}Conclusions}
This Part II concludes the study (Ref.~\cite{Akhm10}) of conditions of efficient heating of a thin conducting cylinder by a broad electromagnetic beam in the longitudinal and transverse geometry.

As a spectacular application, efficient heating of a nanotube by a femtosecond laser pulse is discussed.

\begin{acknowledgments}
The author is very grateful to Dr. A.V. Gavrilin and Dr. A.P. Tarasevitch for valuable remarks and help.
\end{acknowledgments}
\bibliography{dfcdb1}

\end{document}